\documentclass[preprint,aps,superscriptaddress,nofootinbib,tightenlines,floatfix]{revtex4}
\usepackage{subfig}
\usepackage{epsfig}
\usepackage{graphicx}
\usepackage{bm}
\usepackage{rotating}
\usepackage{amsmath}
\usepackage{caption}

\newcommand{\beq}{\begin{equation}}
\newcommand{\eeq}{\end{equation}}
\newcommand{\bea}{\begin{eqnarray}}
\newcommand{\eea}{\end{eqnarray}}

\def\OMIT#1{{}}

\begin{document}
\begin{figure}[!t]
\vskip -1.5cm
\leftline{
{\epsfxsize=1.5in \epsfbox{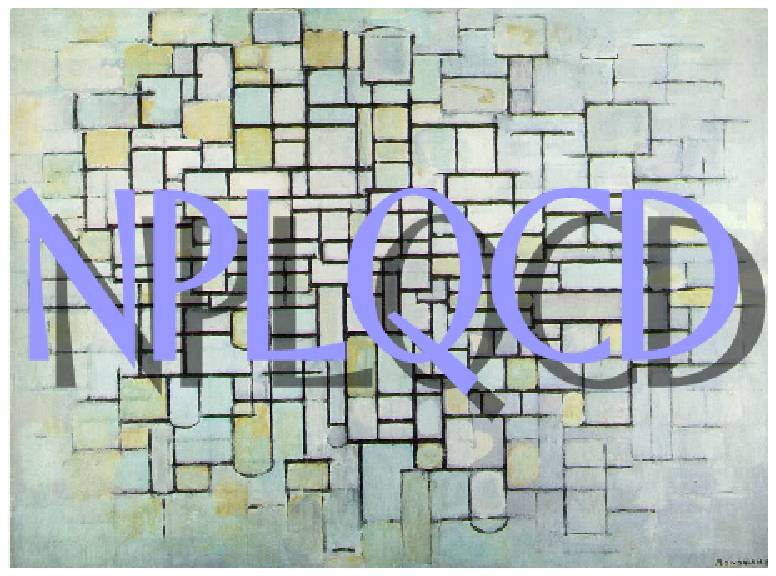}}}
\vskip 1.2cm
\end{figure}

\preprint{\vbox{
\hbox{UNH-09-03}
\hbox{JLAB-THY-09-1021}
\hbox{ICCUB-09-217}
\hbox{ATHENA-PUB-09-017}
\hbox{NT@UW-09-16}
}}

\vskip .5cm

\title{Meson-Baryon Scattering Lengths from \\ Mixed-Action Lattice QCD}

\vskip .5cm
\author{A.~Torok}
\affiliation{Department of Physics, University of New Hampshire,
Durham, NH 03824-3568.}
\author{S.R.~Beane}
\affiliation{Department of Physics, University of New Hampshire,
Durham, NH 03824-3568.}
\author{W.~Detmold}
\affiliation{Department of Physics, College of William and Mary, Williamsburg,
  VA 23187-8795.}
\affiliation{Jefferson Laboratory, 12000 Jefferson Avenue, 
Newport News, VA 23606.}
\author{T.C.~Luu}
\affiliation{N Division, Lawrence Livermore National Laboratory, Livermore, CA 94551.}
\author{K.~Orginos}
\affiliation{Department of Physics, College of William and Mary, Williamsburg,
  VA 23187-8795.}
\affiliation{Jefferson Laboratory, 12000 Jefferson Avenue, 
Newport News, VA 23606.}
\author{A.~Parre\~no}
\affiliation{Departament d'Estructura i Constituents de la Mat\`{e}ria and
Institut de Ci\`encies del Cosmos, 
Universitat de Barcelona,  E--08028 Barcelona, Spain.}
\author{M.J.~Savage}
\affiliation{Department of Physics, University of Washington, 
Seattle, WA 98195-1560.}
\author{A.~Walker-Loud}
\affiliation{Department of Physics, College of William and Mary, Williamsburg,
  VA 23187-8795.}
\collaboration{ NPLQCD Collaboration }
\noaffiliation
\vphantom{}
\vskip 0.8cm

\graphicspath{{figs/}}


\begin{abstract}
The $\pi^+\Sigma^+$, $\pi^+\Xi^0$, $K^+p$, $K^+n$, and
$\overline{K}{}^0 \Xi^0$ scattering lengths are calculated in
mixed-action Lattice QCD with domain-wall valence quarks on the
asqtad-improved coarse MILC configurations at four light-quark masses,
and at two light-quark masses on the fine MILC configurations.  Heavy
Baryon Chiral Perturbation Theory with two and three flavors of light
quarks is used to perform the chiral extrapolations.  To the order we
work in the three-flavor chiral expansion, the kaon-baryon processes
that we investigate show no signs of convergence.  Using the
two-flavor chiral expansion for extrapolation, the pion-hyperon
scattering lengths are found to be $a_{\pi^+\Sigma^+}=-0.197\pm0.017$
fm, and $a_{\pi^+\Xi^0}=-0.098\pm0.017$ fm, where the comprehensive
error includes statistical and systematic uncertainties.
\end{abstract}
\pacs{}
\maketitle



\vfill\eject
\section{Introduction}

\noindent Lattice QCD calculations of meson-meson interactions have
yielded predictions for physical scattering lengths at the few percent
level~\cite{Beane:2007xs,Beane:2006gj,Beane:2007uh}. Several reasons
underlie this striking accuracy. Firstly, at the level of the lattice
calculation, Euclidean-space correlation functions involving
pseudoscalar mesons have signal/noise ratios\footnote{Here the signal
is the Monte Carlo estimate of the quantum correlation function evaluated
on the lattice, while the noise represents the statistical fluctuations in
the correlation function.} that do not degrade, or only slowly
degrade with time. Therefore, highly accurate fits of both single-
and multi-meson properties are possible with currently available
supercomputer resources. Recent calculations of multi-meson
interactions relevant for the study of pion and kaon condensation have been
performed with up to twelve mesons interacting on a
lattice~\cite{Beane:2007es,Detmold:2008fn,Detmold:2008yn} with no
appreciable degradation of signal/noise with time. Secondly, and perhaps
more importantly, QCD correlation functions involving Goldstone bosons
are subject to powerful chiral symmetry constraints. Since current lattice calculations are carried out at unphysical quark masses, these constraints play an essential role in extrapolating the lattice data to the
physical quark masses, as well as to the infinite volume, and continuum
limits. Chiral perturbation theory ($\chi$-PT) is the optimal method
for implementing QCD constraints due to chiral symmetry, and in essence,
provides an expansion of low-energy S-matrix elements in quark masses
and powers of momentum~\cite{Bernard:2007zu}.

In contrast to the purely mesonic sector, recent studies of
baryon-baryon interactions, the paradigmatic nuclear physics process,
have demonstrated the fundamental difficulty faced in making
predictions for baryons and their
interactions~\cite{Beane:2006mx,Beane:2006gf}. Unlike the mesons,
correlation functions involving baryons suffer an exponential
degradation of signal/noise at large times~\footnote{A recent high-statistics study of
  baryon correlation functions on anisotropic clover lattices has
  found that the exponential decay with time of signal/noise occurs only
  {\it asymptotically} in time, and therefore, the signal/noise problem
  in baryon correlation functions is not nearly as severe as
  previously thought~\cite{Beane:2009ky}.} and therefore pose a
fundamentally different kind of challenge in extracting signal from
data~\cite{Lepage:1989hd}. Furthermore, while baryon interactions are
constrained by QCD symmetries like chiral symmetry, the constraints
are not nearly as powerful as when there is at least one pion or kaon
in the initial or final state. For instance, there is no expectation
that the baryon-baryon scattering lengths vanish in the chiral
limit as they do in the purely mesonic sector. In
nucleon-nucleon scattering, the s-wave interactions are actually enhanced due
to the close proximity of a non-trivial fixed point of the
renormalization group, which drives the scattering lengths to
infinity, thus rendering the effective field theory description of the
interaction highly non-perturbative~\cite{Kaplan:1998we}.

Given the contrast in difficulty between the purely mesonic and purely
baryonic sectors described above, it is clearly of great interest to
perform a lattice QCD investigation of the simplest scattering process
involving at least one baryon: meson-baryon scattering. While
pion-nucleon scattering is the best-studied process, both
theoretically and experimentally, its determination on the lattice
is computationally prohibitive since it involves
annihilation diagrams. At present only a few limiting cases that  
involve these diagrams are being investigated~\cite{Babich:2009rq}. Combining the lowest-lying $SU(3)$ meson
and baryon octets, one can form five meson-baryon elastic scattering
processes that do not involve annihilation diagrams. Three of these
involve kaons and therefore are, in principle, amenable to an $SU(3)$
heavy-baryon $\chi$-PT (HB$\chi$-PT) analysis~\cite{Jenkins:1990jv}
for extrapolation. The remaining two processes involve pions
interacting with hyperons and therefore can be analyzed in conjunction
with the kaon processes in $SU(3)$ HB$\chi$-PT, or independently using
$SU(2)$ HB$\chi$-PT.

Meson-baryon scattering has been developed to several non-trivial
orders in the $SU(3)$ HB$\chi$-PT expansion in
Refs.~\cite{Liu:2006xja,Liu:2007ct}, extending earlier work on
kaon-nucleon scattering in Ref.~\cite{Kaiser:2001hr}.  A very-recent
paper~\cite{Mai:2009ce} has reconsidered the $SU(3)$ HB$\chi$-PT
results using a different regularization scheme, and also derived
results for pion-hyperon scattering in the $SU(2)$ HB$\chi$-PT
expansion. These works make clear that the paucity of
experimental data make it is very difficult to assess the convergence of
the chiral expansion in the three-flavor case. Further, in the pion-hyperon system, the  
complete lack of experimental data precludes a separate analysis in  
the chiral two-flavor expansion. A lattice calculation of meson-baryon
scattering analyzed using $\chi$-PT is therefore useful not only in
making predictions for low-energy scattering at the physical point,
but also for assessing the convergence of the chiral expansion for a
range of quark masses at which present-day lattice calculations are
being performed. 

Meson-baryon scattering is also of interest for several indirect
reasons. The $K^- n$ interaction is important for the description of
kaon condensation in the interior of neutron
stars~\cite{KaplanNelson}, and meson-baryon interactions are essential
input in determining the final-state interactions of various decays
that are interesting for standard-model phenomenology (See
Ref.~\cite{Lu:1994ex} for an example). Finally, in determining baryon
excited states on the lattice, it is clear that the energy levels that
represent meson-baryon scattering on the finite-volume lattice must be
resolved before progress can be made regarding the extraction of
single-particle excitations.

The experimental input to existing $\chi$-PT analyses of meson-baryon
scattering is extensively discussed in
Refs.~\cite{Kaiser:2001hr,Liu:2006xja,Liu:2007ct,Mai:2009ce}.
Threshold pion-nucleon scattering information is taken from
experiments with pionic hydrogen and
deuterium~\cite{Schroder:1999uq,Schroder:2001rc}, and the kaon-nucleon
scattering lengths are taken from model-dependent extractions from
kaon-nucleon scattering data~\cite{Martin:1980qe}. There is
essentially no experimental information available on the pion-hyperon
and kaon-hyperon scattering lengths.  There have been two quenched
lattice QCD studies of meson-baryon scattering parameters: the
pioneering work of Ref.~\cite{Fukugita:1994ve} calculated pion-nucleon
and kaon-nucleon scattering lengths at heavy pion masses without
any serious attempt to extrapolate to the physical point, and 
Ref.~\cite{Meng:2003gm} calculated the $I=1$ $KN$ scattering length
and found a result consistent with the current algebra prediction.

In this work we calculate the lowest-lying energy levels for five meson-baryon processes that have no annihilation diagrams:
$\pi^+\Sigma^+$, $\pi^+\Xi^0$, $K^+p$, $K^+n$, and $\overline{K}{}^0 \Xi^0$ in a mixed-action Lattice QCD
calculation with domain-wall valence quarks on the asqtad-improved
coarse MILC configurations with $b\sim 0.125~{\rm fm}$ at four
light-quark masses ($m_\pi\sim 291$, $352$, $491$ and $591$ MeV), and
at two light quark masses ($m_\pi\sim 320$ and $441$ MeV) on the fine
MILC configurations with $b\sim 0.09~{\rm fm}$, with substantially
less statistics on the fine ensembles. We extract the s-wave
scattering lengths from the two-particle energies, and analyze the
five processes using $SU(3)$ HB$\chi$-PT.  We find a rather conclusive
lack of convergence in the three-flavor chiral expansion.  We then
consider $\pi^+\Sigma^+$ and $\pi^+\Xi^0$ using $SU(2)$ HB$\chi$-PT
and find that we are able to make reliable predictions of the
scattering lengths at the physical point. We find
\begin{eqnarray}
a_{\pi^+\Sigma^+}&=& -0.197 \pm 0.017~{\rm fm} \ ;\\
a_{\pi^+\Xi^0}&=& -0.098\pm 0.017~{\rm fm} \ ,
\label{eq:MP}
\end{eqnarray}
where the errors encompass statistical and systematic uncertainties.
The leading order $\chi$-PT (current algebra) predictions for
the scattering lengths are given by~\cite{Weinberg:1966kf}:
\begin{eqnarray}
a_{\pi^+\Sigma^+}&=& -0.2294~{\rm fm} \ ;\\
a_{\pi^+\Xi^0}&=& -0.1158~{\rm fm} \ .
\label{eq:CA}
\end{eqnarray}

Ultimately, either the chiral extrapolation should be performed after
a continuum limit has been taken, or one should use the mixed-action
extension of HB$\chi$-PT to perform the chiral extrapolations~
\cite{Tiburzi:2005is,Chen:2007ug}.  However, our results on the fine MILC configurations are statistics-limited and not yet
sufficiently accurate to make this a useful exercise.  Further, the
explicit extrapolation formulas for the meson-baryon scattering
lengths have not yet been determined in mixed-action $\chi$-PT.
Despite these limitations, we expect the corrections from finite
lattice spacing to be small for two principle reasons.  Firstly, the
meson-baryon scattering lengths are protected by chiral symmetry and
therefore the (approximate) chiral symmetry of the domain wall valence
fermions used in this work protects the scattering lengths from
additive renormalization, which can be explicitly seen in the
construction of the mixed-action baryon Lagrangian in
Ref.~\cite{Chen:2007ug}. The mixed-action corrections do not appear until next-to-next-to leading order in the
chiral expansion of the meson-baryon scattering lengths.  Secondly, our
previous experience with this mixed-action lattice QCD program leads
us to expect that discretization effects will be well-encompassed
within the overall errors we quote.  In our 
precise calculation of meson-meson scattering, the predicted mixed-action
corrections~ \cite{Chen:2005ab,Chen:2006wf} were smaller than the
uncertainties on a given ensemble~\cite{Beane:2007xs,Beane:2007uh}.

This paper is organized as follows. In section~\ref{sec:MBSP} we
isolate the five meson-baryon processes with no annihilation
diagrams that are calculated in this work. We briefly review
the standard L\"uscher method for extracting the scattering amplitude
from two-particle energy levels in a finite volume in section~\ref{sec:finvol}.
Particulars regarding the mixed-action lattice calculation and fitting
methods are provided in section~\ref{sec:MAdetails}.
Additional details can be found in Ref.~\cite{Beane:2008dv}.
Mixing between two of the meson-baryon channels with the same quantum numbers
is discussed in section~\ref{sec:MChAm}.
In section~\ref{sec:su3CE} we consider chiral extrapolations of the
lattice data using $SU(3)$ HB$\chi$-PT, and in section~\ref{sec:su2CE} we
analyze the pion-hyperon lattice data using $SU(2)$ HB$\chi$-PT. Finally,
we conclude in section~\ref{sec:conc}.

\section{Meson-Baryon Scattering Processes}
\label{sec:MBSP}

\noindent It is a straightforward exercise to construct the six scattering channels involving the lowest-lying octet mesons
and baryons that do not have annihilation diagrams, and to determine
their isospin.~\footnote{The $\pi^+\Xi^0$ and $\overline{K}{}^0\Sigma^+$ systems have the same quantum numbers, and therefore require a mixed channel analysis in order to extract the $\overline{K}{}^0\Sigma^+$ scattering length. This is discussed in Section~\ref{sec:MChAm}.} The particle content, isospin, and valence quark
content of these meson-baryon states are shown in Table~\ref{tab:quarks1}.
\begin{table}
\begin{tabular}{ccc}
\hline 
\hline 
Particles\ \ \  &\ \ \   Isospin\ \ \  &\ \ \ Quark Content \\
\hline 
$\pi^+\Sigma^+$ & 2 & $uuu\bar{d}s$  \\
$\pi^+\Xi^0$ & 3/2 & $uu\bar{d}ss$ \\
$K^+p$ & 1 & $uuud\bar{s}$ \\
$K^+n$ & 0~{\it and}~1 & $uudd\bar{s}$ \\
$\overline{K}{}^0\Sigma^+$ & 3/2 & $uu\bar{d}ss$ \\
$\overline{K}{}^0\Xi^0$ & 1 & $u\bar{d}sss$ \\
\hline 
\hline
\end{tabular}
\caption{Particle content, isospin, and valence quark structure of the meson-baryon states calculated in this work.
As is clear from the valence quark content, these meson-baryon states have no annihilation diagrams.}
\label{tab:quarks1}
\end{table}
We adopt the notation of Ref.~\cite{Liu:2006xja}, denoting the threshold T-matrix
in the isospin basis as $T^{(I)}_{\phi B}$, where $I$ is the isospin of the meson-baryon combination, $\phi$ is
the meson, and $B$ is the baryon. The five elastic meson-baryon scattering processes that we consider
are then in correspondence with the isospin amplitudes according to
\begin{eqnarray}
T_{\pi^+\Sigma^+}=T^{(2)}_{\pi \Sigma}\ &;& \qquad T_{\pi^+\Xi^0}=T^{(3/2)}_{\pi \Xi} \ ; \nonumber \\
T_{K^+p}=T^{(1)}_{KN} \ ; \qquad T_{K^+n}&=&\frac{1}{2}(T^{(1)}_{KN}+T^{(0)}_{KN}) \ ; \qquad  T_{\overline{K}{}^0\Xi^0}=T^{(1)}_{\overline{K}\Xi} \ . \nonumber\\ 
\label{eq:Tmatrices}
\end{eqnarray}
These threshold T-matrices are related to the scattering lengths $a_{\phi B}$ through
\begin{equation}
T_{\phi B}=4\pi\left(1+\frac{m_\phi}{m_B}\right) a_{\phi B} \ ,
\label{eq:Tanda}
\end{equation}
where $m_\phi$ is the meson mass and $m_B$ is the baryon mass.

\section{Finite-Volume Calculation of Scattering Amplitudes}
\label{sec:finvol}

\noindent The s-wave scattering amplitude for two particles below
inelastic thresholds can be determined using L\"uscher's
method~\cite{luscher_formula}, which entails a measurement of one or
more energy levels of the two-particle system in a finite volume.  For
two particles with masses $m_\phi$ and $m_B$ in an s-wave, with zero
total three momentum, and in a finite volume, the difference between
the energy levels and those of two non-interacting particles can be
related to the inverse scattering amplitude via the eigenvalue
equation~\cite{luscher_formula}
\begin{eqnarray}
p\cot\delta(p) \ =\ \frac{1}{\pi L}\ {\bf S}\left(\,\frac{p L}{2\pi}\,\right)\ \ ,
\label{eq:energies}
\end{eqnarray}
where $\delta(p)$ is the elastic-scattering phase shift, and
the regulated three-dimensional sum is
\begin{eqnarray}
{\bf S}\left(\,{\eta}\, \right)\ \equiv \ \sum_{\bf j}^{ |{\bf j}|<\Lambda}
\frac{1}{|{\bf j}|^2-\eta^2}\ -\  {4 \pi \Lambda}
\ \ \  .
\label{eq:Sdefined}
\end{eqnarray}
The sum in Eq.~(\ref{eq:Sdefined}) is over all triplets of integers
${\bf j}$ such that $|{\bf j}| < \Lambda$ and the limit
$\Lambda\rightarrow\infty$ is implicit~\cite{Beane:2003da}.  This
definition is equivalent to the analytic continuation of
zeta-functions presented by L\"uscher~\cite{luscher_formula}.  In
Eq.~(\ref{eq:energies}), $L$ is the length of the spatial dimension in
a cubically-symmetric lattice.  The energy eigenvalue, $E_n$, and its
deviation from the sum of the rest masses of the particle, $\Delta
E_n$, are related to the center-of-mass momentum $p_n$, a solution of
Eq.~(\ref{eq:energies}), by
\begin{eqnarray}
\Delta E_n \ & \equiv & E_n\ -\  m_\phi \ - \ m_B \ =\ \sqrt{\ p_n^2\ +\ m_\phi^2\ } \ +\
\sqrt{\ p_n^2\ +\ m_B^2\ }
\ -\ m_\phi\  - \ m_B \ ;
\nonumber\\
& = &  \frac{p_n^2}{2 \mu_{\phi B}}\ +\ ...
\ \ \ ,
\label{eq:energieshift}
\end{eqnarray}
where $\mu_{\phi B}$ is the reduced mass of the meson-baryon system.  In the absence of
interactions between the particles, $|p\cot\delta|=\infty$, and the
energy levels occur at momenta ${\bf p} =2\pi{\bf j}/L$, corresponding
to single-particle modes in a cubic cavity with periodic boundary conditions.  Expanding
Eq.~(\ref{eq:energies}) about zero momenta, $p\sim 0$, one obtains the
familiar relation~\footnote{In order to be consistent with the meson-baryon
literature, we have chosen to use the ``particle
  physics'' definition of the scattering length, as opposed to the
  ``nuclear physics'' definition, which is opposite in sign.}
\begin{eqnarray}
\Delta E_0 &  = &  -\frac{2\pi a}{\mu_{\phi B}  L^3}
\left[\ 1\ +\  c_1 \frac{a}{L}\ +\  c_2 \left( \frac{a}{L} \right)^2 \ \right ]
\ +\ {\cal O}\left(\frac{1}{L^6}\right)
\ \ ,
\label{eq:luscher_a}
\end{eqnarray}
with 
\begin{eqnarray}
c_1 & = & \frac{1}{\pi}
\sum_{{\bf j}\ne {\bf 0}}^{ |{\bf j}|<\Lambda}
\frac{1}{|{\bf j}|^2}\ -\   4 \Lambda \
\ =\ -2.837297
\ \ \ ,\ \ \
c_2\ =\ c_1^2 \ -\ \frac{1}{\pi^2} \sum_{{\bf j}\ne {\bf 0}}
\frac{1}{|{\bf j}|^4}
\ =\ 6.375183
\ ,
\end{eqnarray}
and $a$ is the scattering length, defined by
\begin{eqnarray}
a & = & \lim_{p\rightarrow 0}\frac{\tan\delta(p)}{p} \ .
\label{eq:scatt}
\end{eqnarray}
As the finite-volume lattice calculation cannot achieve $p=0$ (except in the absence of interactions), in quoting
a lattice value for the scattering length extracted from the ground-state
energy level, it is important to determine the error associated with 
higher-order range corrections.

\section{Lattice Calculation and Data Analysis}
\label{sec:MAdetails}

\noindent In calculating the meson-baryon scattering lengths, the
mixed-action lattice QCD scheme was used in which domain-wall
quark~\cite{Kaplan:1992bt,Shamir:1992im,Shamir:1993zy,Shamir:1998ww,Furman:1994ky}
propagators are generated from a smeared source on $n_f = 2+1$
asqtad-improved~\cite{Orginos:1999cr,Orginos:1998ue} rooted, staggered
sea quarks~\cite{Bernard:2001av}.  To improve the chiral symmetry
properties of the domain-wall quarks, hypercubic-smearing
(HYP-smearing)~\cite{Hasenfratz:2001hp,DeGrand:2002vu,DeGrand:2003in}
was used in the gauge links of the valence-quark action.  In the
sea-quark sector, there has been significant debate regarding the
validity of taking the fourth root of the staggered fermion
determinant at finite lattice
spacing~\cite{Durr:2004as,Durr:2004ta,Creutz:2006ys,Bernard:2006zw,Bernard:2006vv,Creutz:2007nv,Bernard:2006ee,Bernard:2006qt,Creutz:2007yg,Creutz:2007pr,Durr:2006ze,Hasenfratz:2006nw,Shamir:2006nj,Sharpe:2006re}.
While there is no proof, there are arguments to suggest that taking
the fourth root of the fermion determinant recovers the contribution
from a single Dirac fermion. The results of this paper assume
that the fourth-root trick recovers the correct continuum limit of
QCD.

The present calculations were performed predominantly with the coarse
MILC lattices with a lattice spacing of $b\sim 0.125$~fm, and a
spatial extent of $L\sim 2.5$~fm.  On these configurations, the
strange quark was held fixed near its physical value while the
degenerate light quarks were varied over a range of masses corresponding to  
the pion masses shown in Table~\ref{tab:MILCcnfs}. See Ref.~\cite{Beane:2008dv} for further details.
Results were also obtained on a coarse MILC ensemble with a spatial
extent of $L\sim 3.5$~fm.  However, this data is statistics
limited. In addition, calculations were performed on two fine MILC
ensembles at $L\sim 2.5$~fm  with $b\sim 0.09$~fm.  On the coarse
MILC lattices, Dirichlet boundary conditions were implemented to
reduce the original time extent of 64 down to 32, which saved a
nominal factor of two in computational time.  While this procedure leads to
minimal degradation of a nucleon signal, it does limit the number of
time slices available for fitting meson properties. By contrast, on
the fine MILC ensembles, anti-periodic boundary conditions were
implemented and all time slices are available.
\begin{table}[!ht]
\begin{ruledtabular}
\begin{tabular}{cccccccc}
Ensemble & $m_\pi$(MeV) & $b m_l$ & $b m_s$ & $b m^{dwf}_l$ & $ b m^{dwf}_s $ & $10^3
\times bm_{res}$~\protect\footnote{Computed by the LHP collaboration for the coarse ensembles.}
& \# of props  \\ 
\hline ({\it i}) 2064f21b676m007m050 &291 & 0.007 & 0.050 & 0.0081 & 0.081 & $1.604\pm 0.038$ & 1039\ $\times$\ 24 
\\ ({\it ii}) 2064f21b676m010m050 &352 & 0.010 & 0.050 & 0.0138 & 0.081 & $1.552\pm 0.027$ & 769\ $\times$\ 24 
\\ ({\it iii}) 2064f21b679m020m050 & 491& 0.020 & 0.050 & 0.0313 & 0.081 & $1.239\pm 0.028$ & 486\ $\times$\ 24 
\\ ({\it iv}) 2064f21b681m030m050 &591 & 0.030 & 0.050 & 0.0478 & 0.081 & $0.982\pm 0.030$ & 564\ $\times$\ 24 
\\ \hline ({\it v}) 2864f21b676m010m050 &352 & 0.010 & 0.050 & 0.0138 & 0.081 & $1.552\pm 0.027$ & 128\ $\times$\ 8 
\\ \hline ({\it vi}) 2896f21b709m0062m031 & 320& 0.0062 & 0.031 & 0.0080 & 0.0423 & $0.380\pm 0.006$ & 1001\ $\times$\ 8 
\\ ({\it vii}) 2896f21b709m0124m031 &441 & 0.0124 & 0.031 & 0.0080 & 0.0423 & $0.380\pm 0.006$ & 513\ $\times$\ 3 
\\
\end{tabular} 
\end{ruledtabular}
\caption{The parameters of the MILC gauge configurations and
  domain-wall propagators used in this work. The subscript $l$ denotes
  light quark (up and down), and $s$ denotes the strange quark. The
  superscript $dwf$ denotes the bare-quark mass for the domain-wall
  fermion propagator calculation. The last column is the number of
  configurations times the number of sources per configuration. 
  Ensembles ({\it i})-({\it iv}) have $L\sim 2.5$~fm and $b\sim 0.125$~fm;
  Ensemble ({\it v}) has $L\sim 3.5$~fm and $b\sim 0.125$~fm;
  Ensembles ({\it vi}),({\it vii}) have $L\sim 2.5$~fm and $b\sim 0.09$~fm.}
\label{tab:MILCcnfs}
\end{table}

The correlation function that projects onto the zero momentum state for
the meson-baryon system is
\begin{equation}
C_{\phi B}(t)={\cal P}_{ij}\sum_{{\bf x,y}}\langle \phi^{\dagger}(t,{\bf x})
\overline{B_i}(t,{\bf y}) \phi(0,{\bf 0}) B_j(0,{\bf 0})\rangle \ ,
\end{equation}
where ${\cal P}_{ij}$ is a positive-energy projector. For instance, in the case of $K^+ p$, the interpolating operators for the
$K^+$ and the proton are
\begin{eqnarray}
\phi(t,{\bf x})&=&K^+(t,{\bf x})=\overline{s}(t,{\bf x})\gamma_5 u(t,{\bf x}) \ ; \nonumber \\
B_i(t,{\bf x})&=&p_i(t,{\bf x})=\epsilon_{abc}u_i^a(t,{\bf x})\left( u^{b\mathrm{T}}(t,{\bf
x})C\gamma_5 d^c(t,{\bf x})\right) \ .
\end{eqnarray}
The masses of the mesons and baryons are extracted using the assumed
form of the large-time behavior of the single particle correlators as
a function of time. As $t\rightarrow \infty$, the ground state
dominates; however, fluctuations of the correlator increase with
respect to the ground state. The meson and baryon two-point
correlators, $C_{\phi}(t)$ and $C_{B}(t)$, behave as
\begin{equation}
C_{\phi}(t) \ \rightarrow \  {\cal A_\mathrm{1}}\ e^{-m_{\phi} \ t}, \qquad C_{B}(t) \ \rightarrow \  {\cal A_\mathrm{2}}\ e^{-m_{B} \ t}\ ,
\label{eq:correlator} 
\end{equation}
respectively, in the limits $t\rightarrow\infty$ and $L\rightarrow\infty$.
In relatively large lattice volumes the energy difference between the
interacting and non-interacting meson-baryon states is a small
fraction of the total energy, which is dominated by the masses of the
mesons and baryons~\cite{Beane:2007xs}.  In order to extract this
energy difference the ratio of correlation functions, $G_{\phi B}(t)$,
is formed
\begin{equation}
G_{\phi B}( t) \equiv 
\frac{C_{\phi B}( t)}{C_{\phi}(t) C_{B}(t)} 
\ = \ \sum_{n=0}^\infty\ {\cal D}_n\ e^{-\Delta E_n\ t}  \ ,
\label{eq:ratio_correlator} 
\end{equation}
where $\Delta E \equiv \Delta E_0$ is the desired  
energy shift. With $\Delta E$, and the extracted masses of the meson and baryon, the scattering length 
can be calculated using Eqs.~(\ref{eq:energies}) and~(\ref{eq:energieshift}),
or, if $a<<L$, from Eq.~(\ref{eq:luscher_a}). For the meson-baryon
scattering lengths calculated in this work, the difference between
the exact and perturbative eigen-equations is negligible.

A variety of fitting methods have been used, including standard
chi-square minimization fits to one and two exponentials. Generalized
effective energy plots are particularly useful for analyzing the
lattice data and for estimating systematic errors~\cite{Beane:2009ky}.
These plots are constructed by taking the ratio of the correlators at
times $t$, and $t+n_J$ (where $n_J$ is an integer)
\begin{equation}
m_{\phi,B}^{\mathrm{eff}}=\frac{1}{n_J}\mathrm{log} \left(\frac{C_{\phi,B}(t)}{C_{\phi,B}(t+n_J)}\right), \qquad
\Delta E_{\phi B}^{\mathrm{eff}}=\frac{1}{n_J}\mathrm{log} \left(\frac{G_{\phi B}(t)}{G_{\phi B}(t+n_J)}\right) \ .
\label{eq:effscatteq}
\end{equation}
With $n_J=1$, the standard effective mass and energy
plots are recovered.  Generalized effective masses form a system of linear equations
for each $n_J$ over the time interval where the data is fit. For
instance, if the interval is given by $\Delta t=t_2-t_1$, then there
is one equation for $m^\mathrm{eff}$ at each $t$, for any $n_J$ that fits within $\Delta t$. The equations
can be solved for $m^\mathrm{eff}$ by casting them into the form of
the so-called normal equation~\cite{Dahl}. Since each $n_J$
constitutes a different effective mass plot, the number of degrees of
freedom is increased significantly. This method provides a fitting
routine that is faster than standard least-squares fitting. Additional
details regarding the utility of generalized effective mass and energy
plots can be found in Ref.~\cite{Beane:2009gs}.

\begin{figure}
     \centering
     \includegraphics[width=0.75\textwidth]{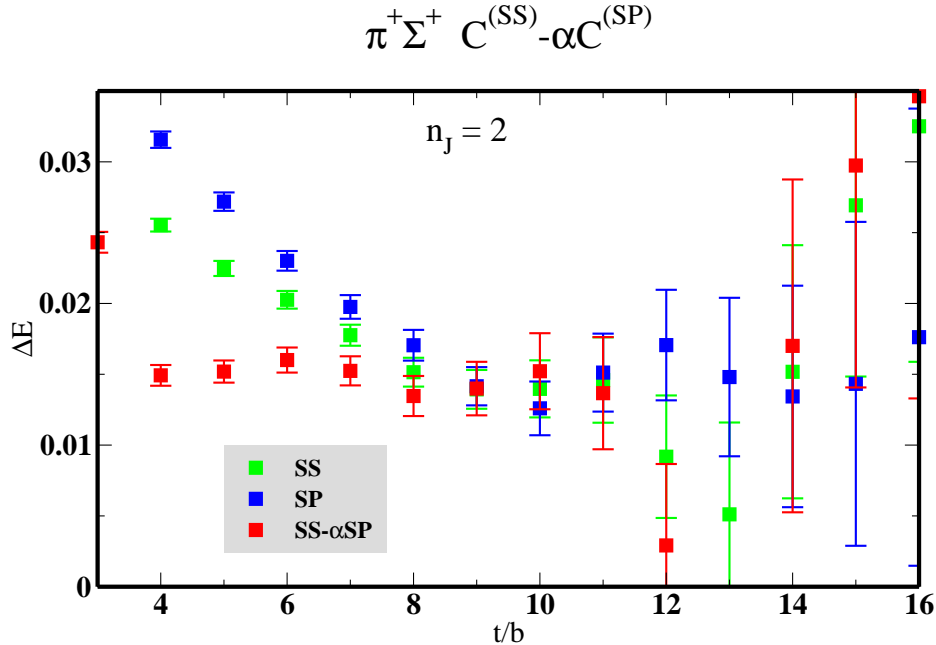}
     \caption{Effective $\Delta E_{\pi^+\Sigma^+}$ plot for coarse
       MILC ensemble ({\it ii}) from correlation functions $C^{\mathrm{(SS)}}$,
       $C^{\mathrm{(SP)}}$ and $C^{\mathrm{(SS)}} \ - \ \alpha C^{\mathrm{(SP)}}$. By taking the linear
       combination with $\alpha$ tuned to remove the first excited
       state, earlier time slices are gained for fitting.}
\label{fig:m010pisigeffSPSS}
\end{figure}

\begin{figure}
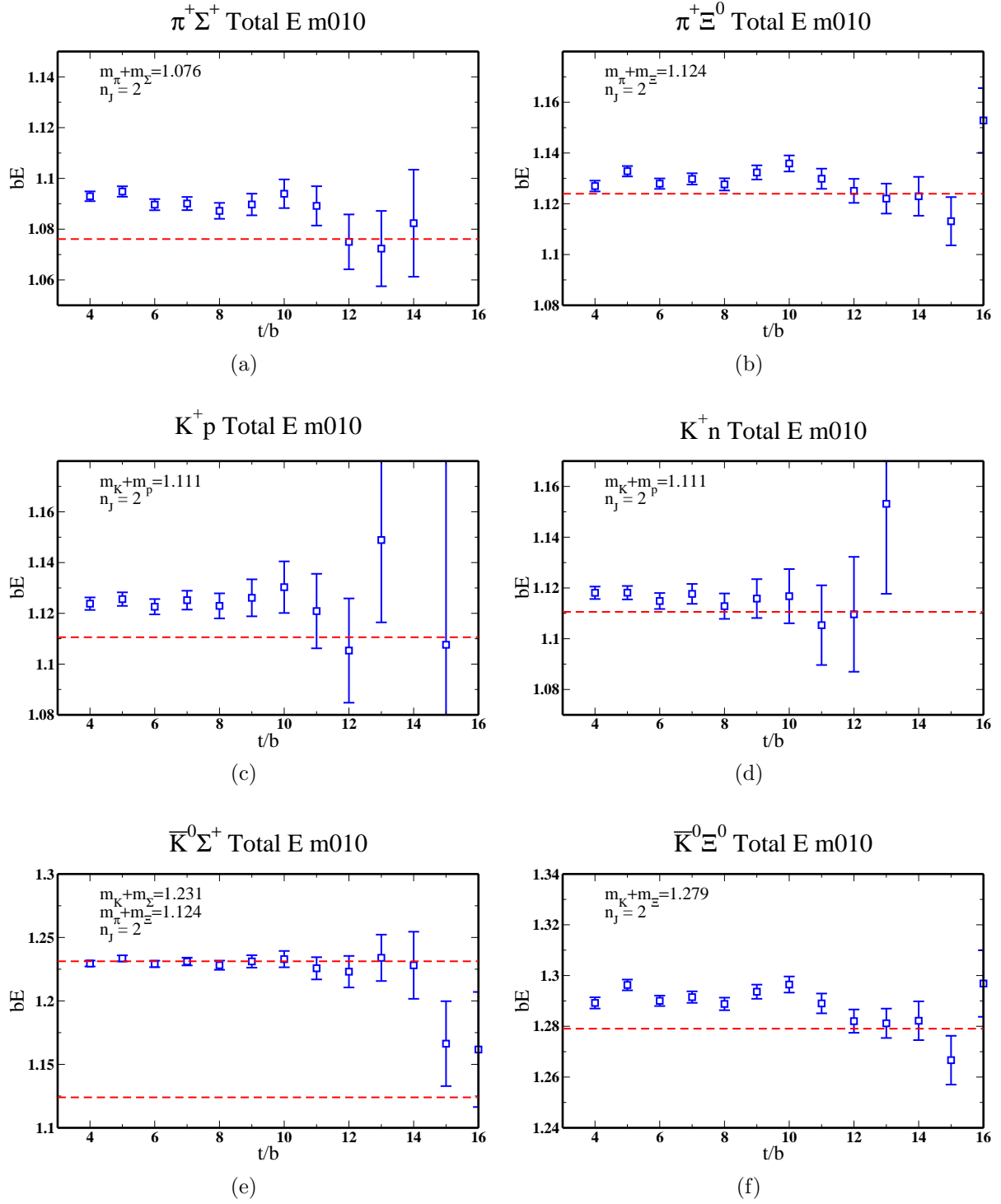

\centering
\subfloat[]{
\includegraphics[width=0.47\linewidth]{Pi_Sigma_E_m010.eps}}
\hspace{1pt}
\vspace{2pt}
\subfloat[]{
\includegraphics[width=0.47\linewidth]{Pi_Xi_E_m010.eps}}
\hspace{1pt}
\vspace{2pt}
\subfloat[]{
\includegraphics[width=0.47\linewidth]{Kaon_Proton_E_m010.eps}}
\hspace{1pt}
\vspace{2pt}
\subfloat[]{
\includegraphics[width=0.47\linewidth]{Kaon_Neutron_E_m010.eps}}
\hspace{1pt}
\vspace{2pt}
\subfloat[]{
\label{fig:energylevels:e}
\includegraphics[width=0.47\linewidth]{Kaon_Sigma_E_m010.eps}}
\hspace{1pt}
\vspace{2pt}
\subfloat[]{
\includegraphics[width=0.47\linewidth]{Kaon_Xi_E_m010.eps}}
\caption{Effective energy plots of the six meson-baryon processes shown in Table~\ref{tab:quarks1}. The plots are from MILC ensemble ({\it ii}), $n_J=2$, and the linear combination $C^{\mathrm{(SS)}} \ - \ \alpha C^{\mathrm{(SP)}}$ is plotted. The dashed line is the sum of the meson and baryon masses for each process, while the error bars represent the jackknife uncertainty. Note that the bE axis of (e) is a factor of two larger in span than the other plots to encompass the dashed line at $m_\pi+m_\Xi=1.124$.}
\label{fig:energylevels}
\end{figure}

\begin{figure}
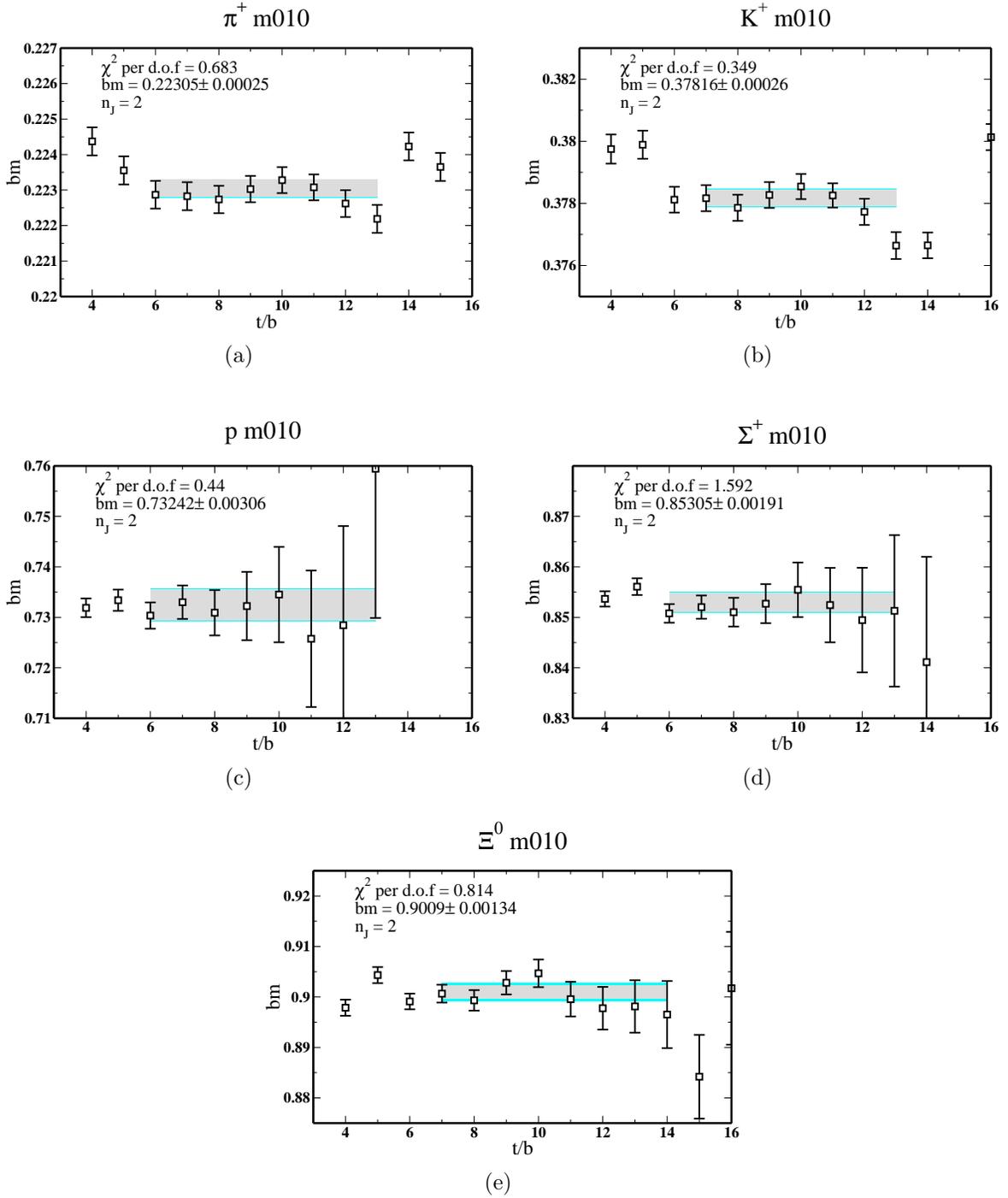

\centering
\subfloat[]{
\label{fig:m010single:a} 
\includegraphics[width=0.45\linewidth]{Pion_m010.eps}}
\hspace{8pt}
\vspace{10pt}
\subfloat[]{
\label{fig:m010single:b} 
\includegraphics[width=0.45\linewidth]{Kaon_m010.eps}}
\hspace{8pt}
\vspace{10pt}
\subfloat[]{
\label{fig:m010single:c} 
\includegraphics[width=0.45\linewidth]{Proton_m010.eps}}
\hspace{8pt}
\subfloat[]{
\label{fig:m010single:d} 
\includegraphics[width=0.45\linewidth]{Sigma_m010.eps}}
\hspace{8pt}
\subfloat[]{
\label{fig:m010single:e} 
\includegraphics[width=0.45\linewidth]{Xi_m010.eps}}
\caption{Single particle effective mass plots for coarse MILC
       ensemble ({\it ii}). Here we choose $n_J=2$, and the linear 
       combination $C^{\mathrm{(SS)}}-\alpha C^{\mathrm{(SP)}}$ is plotted. The inner shaded bands are 
       the jackknife uncertainties of the fits to the effective masses, and the outer bands are the jackknife uncertainty and systematic uncertainty added in quadrature over the indicated window of time slices.}
\label{fig:m010single}
\end{figure}

\begin{figure}
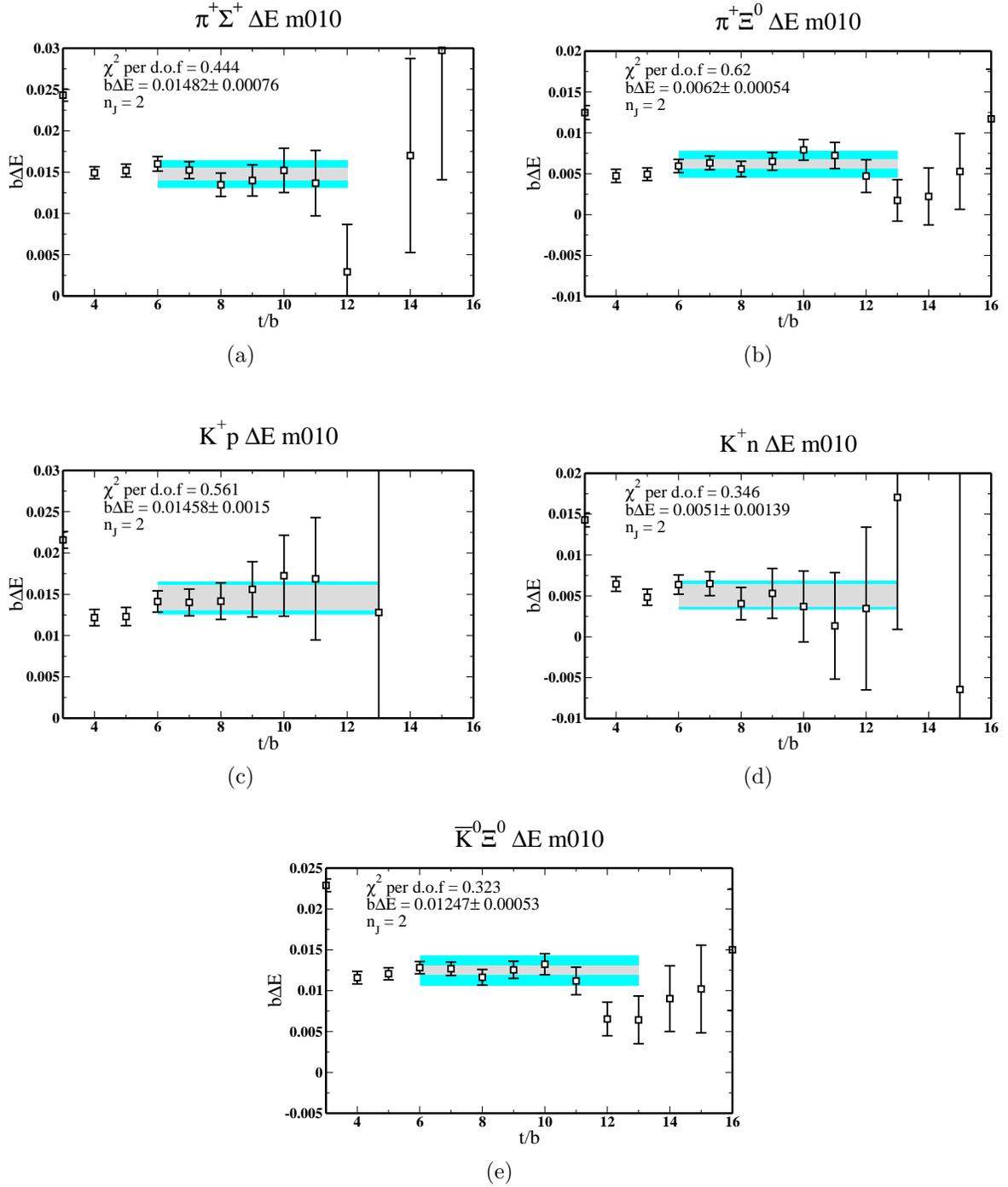

\centering
\subfloat[]{
\label{fig:m010two:f} 
\includegraphics[width=0.45\linewidth]{Pi_Sigma_m010.eps}}
\hspace{8pt}
\vspace{10pt}
\subfloat[]{
\label{fig:m010two:g} 
\includegraphics[width=0.45\linewidth]{Pi_Xi_m010.eps}}
\hspace{8pt}
\vspace{10pt}
\subfloat[]{
\label{fig:m010two:h} 
\includegraphics[width=0.45\linewidth]{Kaon_Proton_m010.eps}}
\hspace{8pt}
\subfloat[]{
\label{fig:m010two:i} 
\includegraphics[width=0.45\linewidth]{Kaon_Neutron_m010.eps}}
\hspace{8pt}
\subfloat[]{
\label{fig:m010two:k} 
\includegraphics[width=0.45\linewidth]{Kaon_Xi_m010.eps}}
\caption{Meson-baryon effective energy difference plots for coarse MILC
       ensemble ({\it ii}). Here we choose $n_J=2$, and the linear 
       combination $C^{\mathrm{(SS)}}-\alpha C^{\mathrm{(SP)}}$ is plotted. The inner shaded bands
       are the jackknife uncertainties of the fits to the effective energy differences, and the outer bands are the jackknife uncertainty and systematic uncertainty added in quadrature over the indicated window of time slices.}
\label{fig:m010two}
\end{figure}

\begin{figure}
     \centering
     \includegraphics[width=0.75\textwidth]{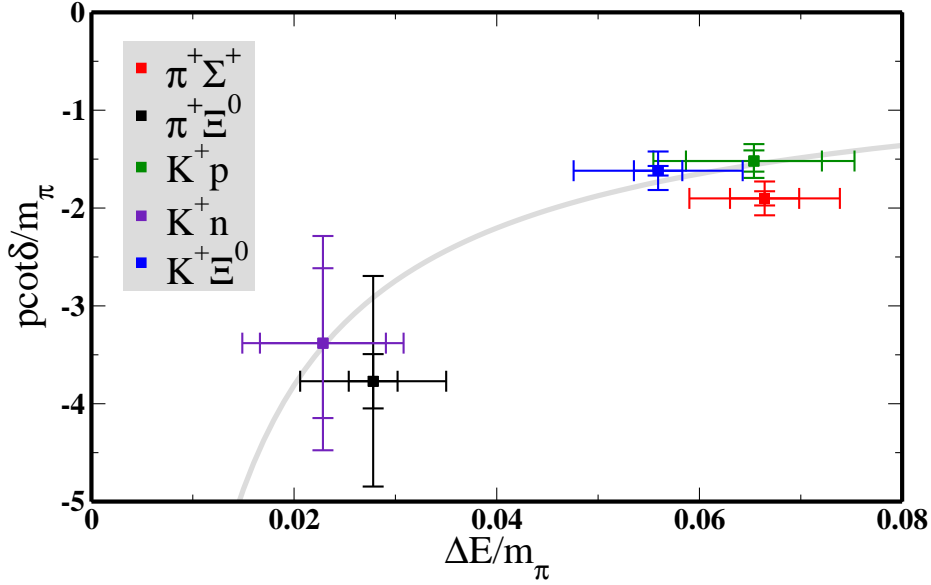}
     \caption{$p\cot\delta/m_\pi$ versus $\Delta E_{\phi B}/m_\pi$ for the five elastic scattering processes from coarse MILC ensemble ({\it ii}). The curve shown is $p\cot\delta/m_\pi$ for the case of $m_\phi=m_K$, and $m_B=m_p$.}
\label{fig:m010Savageplot}
\end{figure}

The interpolating operator at the source is constructed from
gauge-invariantly-smeared quark field operators, while at the sink,
the interpolating operator is constructed from either local quark
field operators, or from the same smeared quark field operators used
at the source, leading to two sets of correlation functions.  For
brevity, we refer to the two sets of correlation functions that result
from these source and sink operators as {\it smeared-point} (SP) and
{\it smeared-smeared} (SS) correlation functions, respectively.  By
forming a linear combination of the SP and SS correlation functions,
$C^{\mathrm{(SS)}} \ - \ \alpha C^{\mathrm{(SP)}}$, we are able to remove
the first excited state, thus gaining early time slices for
fitting~\cite{Beane:2009gs}.  This effect is illustrated in
Fig.~\ref{fig:m010pisigeffSPSS}, which is the effective $\Delta
E_{\pi^+\Sigma^+}$ plot for coarse MILC ensemble ({\it ii}). We plot
$C^{\mathrm{(SS)}}$, $C^{\mathrm{(SP)}}$, and $C^{\mathrm{(SS)}} \ - \ \alpha
C^{\mathrm{(SP)}}$ with $\alpha$ tuned to remove the first excited
state. The effective energies, effective masses, and energy splittings are plotted for
coarse MILC ensemble ({\it ii}) in Figs.~\ref{fig:energylevels},~\ref{fig:m010single}, and
\ref{fig:m010two}. All of the necessary quantities needed for
extraction of the scattering lengths are contained in
Table~\ref{tab:latticequant}, which also contains the sum of meson and
baryon masses at each quark mass. Fig.~\ref{fig:m010Savageplot} shows the results for all five processes, and the behavior of Eq.~(\ref{eq:energies}), versus the interaction energy, presented in terms of the dimensionless quantities $p\cot\delta/m_\pi$ and $\Delta E/m_\pi$. The curve shown in Fig.~\ref{fig:m010Savageplot} is $p\cot\delta/m_\pi$ for the case of $m_\phi=m_K$, and $m_B=m_p$, as $\Delta E/m_\pi$ is varied. ${\bf S}(\eta)$ in Eq.~(\ref{eq:Sdefined}) is a function of the meson and baryon masses, so there will be a unique curve for each combination of $m_\phi$ and $m_B$. Consequently, the $K^+p$, and $K^+n$ data points fall on this curve.

%

\begin{table}
\begin{ruledtabular}
\begin{tabular}{ccccc}
Quantity & m007 ({\it i})& m010 ({\it ii})& m020 ({\it iii})& m030 ({\it iv})  \\
\hline
$m_{\pi}$ &  0.18384(31)(03) &  0.22305(25)(08) &   0.31031(38)(95) &   0.37513(44)(13)  \\
$m_{k}$ &  0.36783(32)(42) &  0.37816(26)(11) &   0.40510(33)(37) &   0.43091(66)(16)  \\
$m_{p}$ &  0.6978(61)(08) &  0.7324(31)(10) &   0.8069(22)(14) &   0.8741(16)(05)  \\
$m_{\Sigma}$ &  0.8390(22)(03) &  0.8531(19)(08) &   0.8830(18)(17) &   0.9213(13)(03)  \\
$m_{\Xi}$ &  0.8872(13)(16) &  0.9009(13)(10) &   0.9233(18)(04) &   0.9461(14)(08) \\
$f_{\pi}$ &  0.09257(16) &  0.09600(14) &   0.10208(14) &   0.10763(32)  \\
$f_{K}$ &  0.10734(10) &  0.10781(18) &   0.10976(17) &   0.11253(31) \\
\hline
$\Delta E_{\pi\Sigma}$ &  0.0150(14)(08) &  0.0148(08)(13) &   0.0111(10)(08) &   0.0100(10)(11)  \\
$\Delta E_{\pi\Xi}$ &  0.00646(64)(98) &  0.0062(05)(12) &   0.00431(68)(43) &   0.00421(76)(60)  \\
$\Delta E_{K p}$ &  0.0140(22)(30) &  0.0146(15)(13) &   0.0092(10)(51) &   0.0087(16)(16) \\
$\Delta E_{K n}$ &  0.0057(18)(16) &  0.0051(14)(09) &   0.0036(09)(12) &   0.0028(10)(11)  \\
$\Delta E_{K\Xi}$ &  0.0118(08)(13) &  0.0125(05)(14) &   0.0085(08)(31) &   0.0086(16)(16)  \\
\hline
$a_{\pi\Sigma}$ & -2.12(16)(09) & -2.36(09)(15) &  -2.30(15)(13) &  -2.36(18)(19)\\
$a_{\pi\Xi}$ & -1.08(09)(14) & -1.19(09)(20) &  -1.08(15)(09) &  -1.20(18)(15) \\
$a_{Kp}$ & -2.80(32)(44) & -2.95(21)(19) &  -2.3(0.2)(1.0) &  -2.27(31)(32) \\
$a_{Kn}$ & -1.41(37)(34) & -1.33(30)(21) &  -1.05(22)(30) &  -0.89(27)(31)  \\
$a_{K\Xi}$ & -2.62(13)(21) & -2.77(08)(23) &  -2.18(15)(63) &  -2.29(30)(32)  \\
\hline
$m_{\pi}+m_{p}$ &  0.8817(61) &  0.9555(31) &   1.1172(23) &   1.2492(18)\\
$m_{\pi}+m_{\Sigma}$ &  1.0229(23) &  1.0761(20) &   1.1933(19) &   1.2964(15)  \\
$m_{\pi}+m_{\Xi}$ &  1.0710(14) &  1.1240(14) &   1.2336(19) &   1.3212(16)  \\
$m_{K}+m_{p}$ &  1.0657(61) &  1.1106(31) &   1.2119(23) &   1.3050(19)  \\
$m_{K}+m_{\Sigma}$ &  1.2069(23) &  1.2312(20) &   1.2881(19) &   1.3522(16) \\
$m_{K}+m_{\Xi}$ &  1.2550(14) &  1.2791(15) &   1.3284(19) &   1.3770(17) \\
\end{tabular}
\end{ruledtabular}
\caption{Lattice calculation results from the four coarse MILC ensembles which enter the analysis
of the meson-baryon scattering lengths. The first uncertainty is statistical and the second uncertainty is systematic
due to fitting. All quantities are in lattice units.}
\label{tab:latticequant}
\end{table}

\section{The Mixed Channel}
\label{sec:MChAm}

As is clear from Table I, the $\pi^+\Xi^0$ and
$\overline{K}{}^0\Sigma^+$  states carry the same 
global quantum numbers, and therefore couple to the same 
energy-eigenstates in  the finite lattice volume.
For energies above both kinematic thresholds, a determination of 
the three scattering parameters associated with these states
(two phases and one mixing-angle) requires a coupled-channel analysis.
Therefore,  three energy levels above both kinematic thresholds must be 
determined in the lattice  calculation to fully  characterize 
scattering in this kinematic regime. In the present lattice
volumes, the two-particle energies  in these channels are
close to the respective kinematic thresholds, and the 
energy of the lower-lying $\pi^+\Xi^0$ state (which is below the $\overline{K}{}^0\Sigma^+$ threshold)
is determined by the low-energy elastic scattering parameters, 
making it amenable to analysis  using Eqs.~(\ref{eq:energies}),
(\ref{eq:Sdefined}), (\ref{eq:energieshift}) and (\ref{eq:luscher_a}).

A priori, one would expect both the $\pi^+\Xi^0$ and
$\overline{K}{}^0\Sigma^+$ interpolating operators to couple to a common
ground state (dominantly the $\pi^+ \Xi^0$ state), with a
$\overline{K}{}^0\Sigma^+$-related level as the first excited state (for
the lattice volumes considered here, the non-interacting $\pi^+\Xi^0$
system with two units of relative momentum has an energy considerably
above the $\overline{K}{}^0\Sigma^+$ threshold). Interestingly, within
our statistical and systematic uncertainties, we find distinct energy
levels from the two interpolating operators. This is consistent with
strong coupling to the color-singlet constituents of the interpolating
operator and only very weak couplings to states that require color
rearrangement (see Fig.~\ref{fig:energylevels}).  While this is
suggestive that mixing between the states is small, a definitive
interpretation requires an extraction of three energy levels above the kinematic thresholds of the $\pi^+\Xi^0$ and
$\overline{K}{}^0\Sigma^+$, and below the next kinematic threshold, in order to determine the three 
scattering parameters. The optimal way to extract these levels
is to use the variational method~\cite{Michael:1985ne,Luscher:1990ck}, which
requires the full matrix of correlation functions to be calculated,
and diagonalized. The extraction of the scattering parameters would
then proceed via an extension of the variational method to the
coupled-channel scenario~\cite{Detmold:2004qn,He:2005ey}.

Due to our incomplete knowledge of the three mixed-channel energy levels, we
do not attempt to extract any $\overline{K}{}^0\Sigma^+$ scattering
parameters in this work.

\section{SU(3) HB$\chi$PT Extrapolation}
\label{sec:su3CE}

\subsection{Scattering Length Formulas}
\label{sec:scattLextrap}

\noindent The scattering lengths of the five meson-baryon processes
listed in Eq.~(\ref{eq:Tmatrices}) are, to $\mathcal{O}(m_\pi^3)$ in $SU(3)$
HB$\chi$-PT~\cite{Liu:2006xja,Liu:2007ct},
\begin{eqnarray}
a_{\pi^+\Sigma^+}=\frac{1}{4 \pi}\frac{m_\Sigma}{m_\pi+m_\Sigma} \bigg[ -\frac{2m_\pi}{f_\pi^2} + \frac{2m_\pi^2}{f_\pi^2}C_1 + \mathcal{Y}_{\pi^+\Sigma^+}(\mu ) 
+ 8 h_{123}(\mu )\frac{m_\pi^3}{f_\pi^2} \bigg] \ ;
\label{eq:apisigfull}
\end{eqnarray}
\begin{eqnarray}
a_{\pi^+\Xi^0}=\frac{1}{4 \pi}\frac{m_\Xi}{m_\pi+m_\Xi} \bigg[ -\frac{m_\pi}{f_\pi^2} + \frac{m_\pi^2}{f_\pi^2}C_{01} + \mathcal{Y}_{\pi^+\Xi^0}(\mu ) + 8 h_1(\mu )\frac{m_\pi^3}{f_\pi^2} \bigg] \ ;
\label{eq:apixifull}
\end{eqnarray}
\begin{eqnarray}
a_{K^+ p}=\frac{1}{4 \pi}\frac{m_N}{m_K+m_N} \bigg[ -\frac{2m_K}{f_K^2} + \frac{2m_K^2}{f_K^2}C_1 + \mathcal{Y}_{K^+ p}(\mu ) + 8 h_{123}(\mu )\frac{m_K^3}{f_K^2} \bigg] \ ;
\label{eq:akpfull}
\end{eqnarray}
\begin{eqnarray}
a_{K^+ n}=\frac{1}{4 \pi}\frac{m_N}{m_K+m_N} \bigg[ -\frac{m_K}{f_K^2} + \frac{m_K^2}{f_K^2}C_{01} + \mathcal{Y}_{K^+ n}(\mu ) + 8 h_1(\mu )\frac{m_K^3}{f_K^2} \bigg] \ ;
\label{eq:aknfull}
\end{eqnarray}
\begin{eqnarray}
a_{\overline{K}{}^0 \Xi^0}=\frac{1}{4 \pi}\frac{m_\Xi}{m_K+m_\Xi} \bigg[ -\frac{2m_K}{f_K^2} + \frac{2m_K^2}{f_K^2}C_1 + \mathcal{Y}_{\overline{K}{}^0 \Xi^0}(\mu ) 
+ 8 h_{123}(\mu )\frac{m_K^3}{f_K^2} \bigg] \ ,
\label{eq:akxifull}
\end{eqnarray}
where we have defined $C_{01}\equiv C_0+C_1$ and $h_{123}\equiv h_1-h_2+h_3$, and the loop functions are given by
\begin{eqnarray}
\mathcal{Y}_{\pi^+\Sigma^+}(\mu )&=&\frac{m_\pi^2}{2\pi^2
f_\pi^4}\bigg\{-m_\pi\bigg(\frac32-2\ln\frac{m_\pi}{\mu}-\ln\frac{m_K}{\mu}\bigg) \nonumber \\
&& -\sqrt{m_K^2-m_\pi^2}\arccos\frac{m_\pi}{m_K} + \frac{\pi}{2}\bigg[3F^2 m_\pi-\frac13 D^2 m_\eta\bigg]\bigg\} \ ;
\label{eq:pisigloop}
\end{eqnarray}
\begin{eqnarray}
\mathcal{Y}_{\pi^+ \Xi^0}(\mu )&=&\frac{m_\pi^2}{4\pi^2 f_\pi^4} \bigg\{ -m_\pi
\bigg( \frac32 -2\ln\frac{m_\pi}{\mu}-\ln\frac{m_K}{\mu}\bigg) -\sqrt{m_K^2-m_\pi^2}\bigg(\pi +
\arccos\frac{m_\pi}{m_K}\bigg) \nonumber\\
&& +\frac{\pi}{4}\bigg[3(D-F)^2
m_\pi-\frac13(D+3F)^2 m_\eta\bigg]\bigg\} \ ;
\label{eq:pixiloop}
\end{eqnarray}
\begin{eqnarray}
\mathcal{Y}_{K^+p}(\mu )&=&\frac{m_K^2}{4\pi^2 f_K^4}\bigg\{m_K
\bigg(-3+2\ln\frac{m_\pi}{\mu} + \ln\frac{m_K}{\mu}+3
\ln\frac{m_\eta}{\mu} \bigg)  \nonumber \\
&& +2\sqrt{m_K^2-m_\pi^2} \ln\frac{m_K+\sqrt {m_K^2-m_\pi^2}}{m_\pi}
-3\sqrt{m_\eta^2-m_K^2}\arccos\frac{m_K}{m_\eta} \nonumber\\
&& - \frac{\pi}{6} (D-3F)\bigg[ 2(D+F) \frac{m_\pi^2}{m_\eta+m_\pi}
+(D+5F) m_\eta  \bigg] \bigg\} \ ;
\label{eq:kploop}
\end{eqnarray}
\begin{eqnarray}
\mathcal{Y}_{K^+n}(\mu )&=&\frac{\mathcal{Y}_{K^+p}}{2} + \frac{3m_K^2}{8\pi^2 f_K^4}\bigg\{m_K \bigg(
\ln\frac{m_\pi}{\mu}-\ln\frac{m_K}{\mu} \bigg) + \sqrt{m_K^2-m_\pi^2} \ln\frac{m_K+\sqrt{m_K^2-m_\pi^2}}{m_\pi}\nonumber\\
&& + \frac{\pi}{3} (D-3F) \bigg[(D+F) \frac{m_\pi^2}{m_\eta+m_\pi}
+\frac16(7D+3F) m_\eta \bigg] \bigg\} \ ;
\label{eq:knloop}
\end{eqnarray}
\begin{eqnarray}
\mathcal{Y}_{\overline{K}{}^0\Xi^0}^{(1)}(\mu )&=&\frac{m_K^2}{4\pi^2 f_K^4}\bigg\{m_K
\bigg(-3+2\ln\frac{m_\pi}{\mu} + \ln\frac{m_K}{\mu}+3
\ln\frac{m_\eta}{\mu} \bigg)  \nonumber \\
&& +2\sqrt{m_K^2-m_\pi^2} \ln\frac{m_K+\sqrt {m_K^2-m_\pi^2}}{m_\pi}  -3\sqrt{m_\eta^2-m_K^2}\arccos\frac{m_K}{m_\eta} \nonumber\\
&& - \frac{\pi}{6} (D+3F)\bigg[ 2(D-F) \frac{m_\pi^2}{m_\eta+m_\pi}
+(D-5F) m_\eta  \bigg] \bigg\} \ .
\label{eq:kxiloop}
\end{eqnarray}
In what follows, we choose $\mu=\Lambda_\chi=4\pi f_\pi$ and evaluate
$f_\pi$ at its lattice physical value~\cite{Beane:2005rj}, and we take
$m_\eta$ from the Gell-Mann-Okubo formula. These choices modify the
chiral expansion at $\mathcal{O}(m_\pi^4)$ and are therefore consistent
to the order we are working. The first mixed-action modification to these HB$\chi$-PT  
extrapolation formulas appear as corrections to these loop  
functions, $\mathcal{Y}_{\phi B}$, and to the corresponding counterterms  
which absorb the scale dependence.  Some of the mesons propagating in  
the loops appear as mixed valence-sea combinations, and thus the  
corresponding meson masses appearing in these functions are heavier by a known amount~\cite{Orginos:2007tw}.  The precise form of the predicted corrections require a computation of the  
scattering processes with mixed-action/partially quenched $\chi$-PT.

Our physical parameters are consistent with
Ref.~\cite{Mai:2009ce} (note that our decay constant convention
differs by $\sqrt{2}$). Namely, $f_\pi=130.7~{\rm MeV}$,
$m_\pi=139.57~{\rm MeV}$, $f_K=159.8~{\rm MeV}$, $m_K=493.68~{\rm
  MeV}$, $m_N=938~{\rm MeV}$, $m_\Sigma=1192~{\rm MeV}$ and
$m_\Xi=1314~{\rm MeV}$. The axial couplings, $D$ and $F$, for
coarse MILC ensembles ({\it ii})-({\it iv}) are taken from the mixed-action
calculation of Ref.~\cite{Lin:2007ap}, and we extrapolate for
coarse MILC ensemble ({\it i}) using these values.

\subsection{Extrapolation to the Physical Point}

\begin{figure}
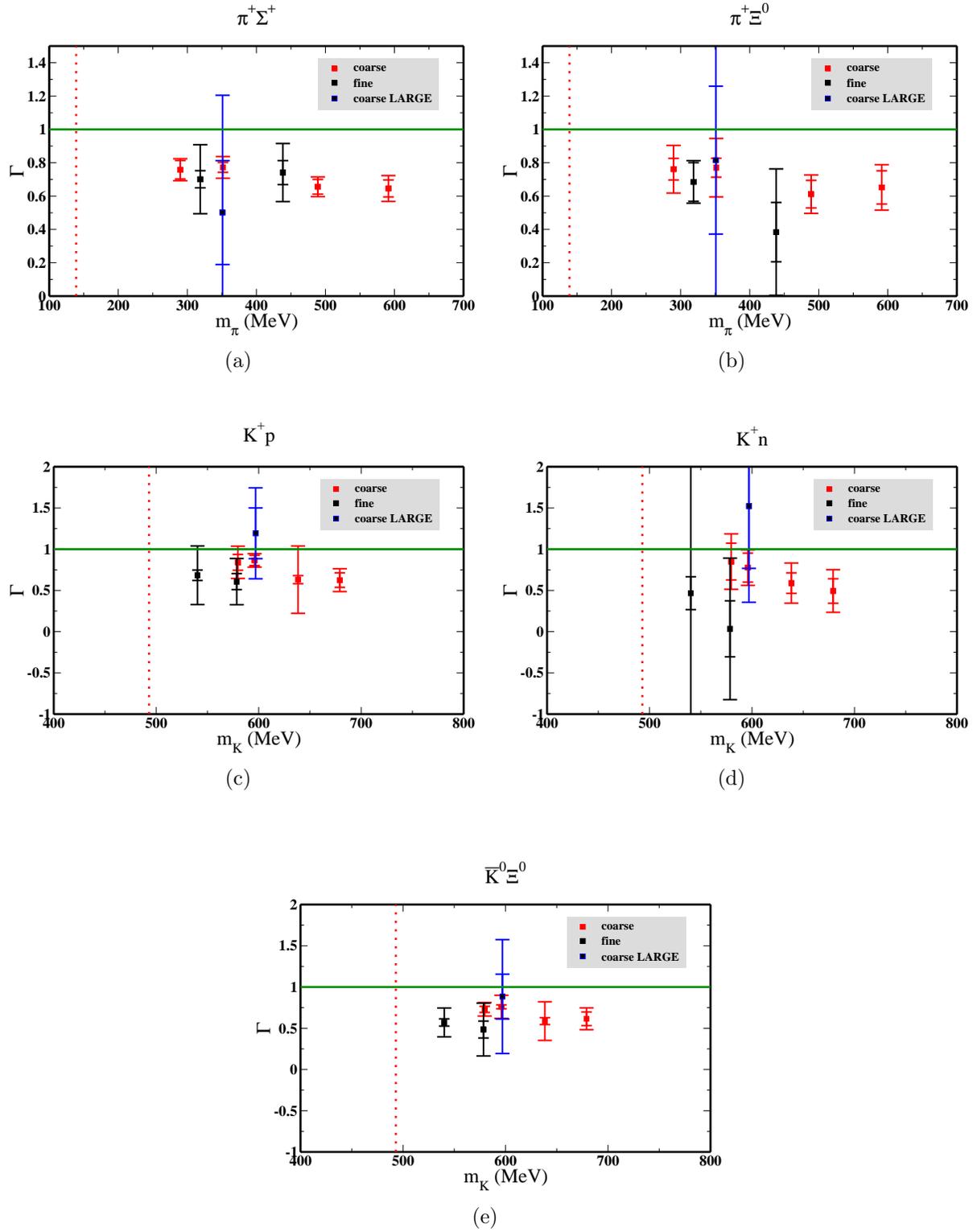

\centering
\subfloat[]{
\includegraphics[width=0.47\linewidth]{Pi_SigmaKostasNLOALL.eps}}
\hspace{1pt}
\vspace{10pt}
\subfloat[]{
\includegraphics[width=0.47\linewidth]{Pi_XiKostasNLOALL.eps}}
\hspace{1pt}
\vspace{10pt}
\subfloat[]{
\includegraphics[width=0.47\linewidth]{Kaon_ProtonKostasNLOALL.eps}}
\hspace{1pt}
\vspace{10pt}
\subfloat[]{
\includegraphics[width=0.47\linewidth]{Kaon_NeutronKostasNLOALL.eps}}
\hspace{1pt}
\vspace{10pt}
\subfloat[]{
\includegraphics[width=0.47\linewidth]{Kaon_XiKostasNLOALL.eps}}
\caption{Plots of $\Gamma_{NLO}$ versus the Goldstone masses for the five meson-baryon processes. All lattice data is included.}
\label{fig:KostasAlldata}
\end{figure}

\noindent For the purposes of fitting and visualization, it is useful to construct
from the scattering lengths the functions $\Gamma^{(1,2)}$ which are
polynomials in $m_\phi$.  For the $\pi^+\Sigma^+$, $K^+p$, and
$\overline{K}{}^0\Xi^0$ processes one defines\footnote{Here we use the standard
notation, LO = leading order, NLO = next-to-leading order and so on.}
\begin{eqnarray}
\Gamma_{LO}^{(1)}\equiv-\frac{2 \pi a f_\phi^2}{m_\phi}\bigg(1 + \frac{m_\phi}{m_B}\bigg)=1 \ ;
\label{eq:GammaLHSLO1}
\end{eqnarray}
\begin{eqnarray}
\Gamma_{NLO}^{(1)}\equiv-\frac{2 \pi a f_\phi^2}{m_\phi}\bigg(1 + \frac{m_\phi}{m_B}\bigg)=1-C_1 m_\phi \ ;
\label{eq:GammaLHSNLO1}
\end{eqnarray}
\begin{eqnarray}
\Gamma_{NNLO}^{(1)}\equiv
-\frac{2 \pi a f_\phi^2}{m_\phi}\bigg(1 + \frac{m_\phi}{m_B}\bigg) + \frac{f_\phi^2}{2m_\phi}\mathcal{Y}_{\phi B}(\Lambda_{\chi})
=1-C_1 m_\phi-4h_{123}(\Lambda_{\chi}) m_\phi^2 \ ,
\label{eq:GammaLHSNNLO1}
\end{eqnarray} 
and for the $\pi^+\Xi^0$, and $K^+n$ processes one defines
\begin{eqnarray}
\Gamma_{LO}^{(2)}\equiv-\frac{4 \pi a f_\phi^2}{m_\phi}\bigg(1 + \frac{m_\phi}{m_B}\bigg)=1 \ ;
\label{eq:GammaLHSLO2}
\end{eqnarray}
\begin{eqnarray}
\Gamma_{NLO}^{(2)}\equiv-\frac{4 \pi a f_\phi^2}{m_\phi}\bigg(1 + \frac{m_\phi}{m_B}\bigg)=1-C_{01} m_\phi \ ;
\label{eq:GammaLHSNLO2}
\end{eqnarray}
\begin{eqnarray}
\Gamma_{NNLO}^{(2)}\equiv
-\frac{4 \pi a f_\phi^2}{m_\phi}\bigg(1 + \frac{m_\phi}{m_B}\bigg) + \frac{f_\phi^2}{m_\phi}\mathcal{Y}_{\phi B}(\Lambda_{\chi})
=1-C_{01} m_\phi-8h_1(\Lambda_{\chi}) m_\phi^2 \ .
\label{eq:GammaLHSNNLO2}
\end{eqnarray}
Notice that the left-hand sides of these equations are given entirely
in terms of lattice-determined quantities, all evaluated under
Jackknife, whereas the right-hand side provides a convenient
polynomial fitting function. Plots of $\Gamma_{NLO}$ formed from the
lattice data (all ensembles listed in Table~\ref{tab:MILCcnfs}) versus
the Goldstone masses are given in Fig.~\ref{fig:KostasAlldata}.  We
see evidence in this plot that the fine and large-volume coarse data are
statistically limited as compared to the coarse data. Therefore, we
include only the coarse data in our fits. The fine data is, however,
indicative that lattice-spacing effects are small.

\begin{figure}
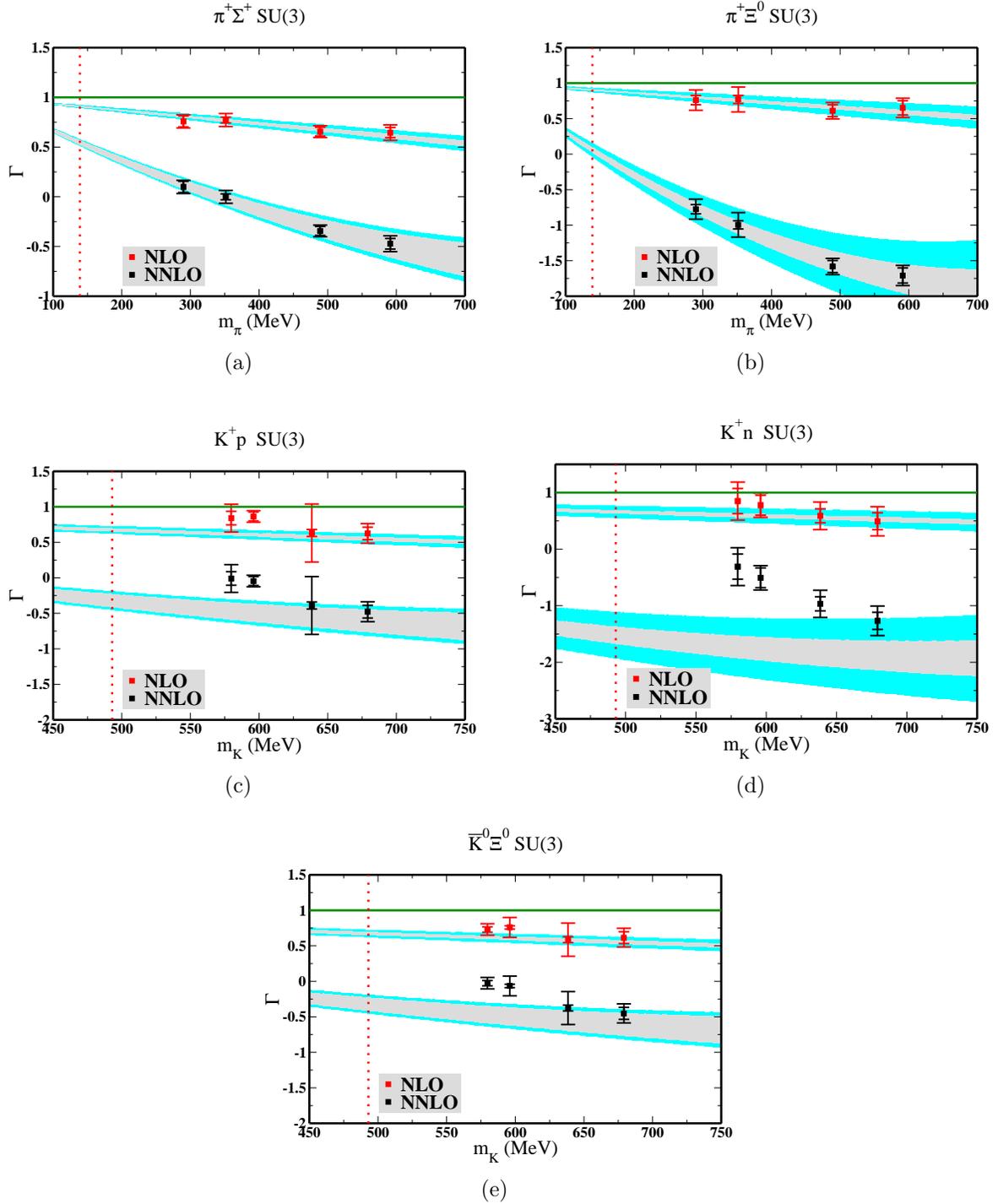

\centering
\subfloat[]{
\label{fig:KostasplotsNLO:a} 
\includegraphics[width=0.45\linewidth]{Pi_Sigma_SU3.eps}}
\hspace{8pt}
\vspace{10pt}
\subfloat[]{
\label{fig:KostasplotsNLO:b} 
\includegraphics[width=0.45\linewidth]{Pi_Xi_SU3.eps}}
\hspace{8pt}
\vspace{10pt}
\subfloat[]{
\label{fig:KostasplotsNLO:c} 
\includegraphics[width=0.45\linewidth]{Kaon_Proton_SU3.eps}}
\hspace{8pt}
\subfloat[]{
\label{fig:KostasplotsNLO:d} 
\includegraphics[width=0.45\linewidth]{Kaon_Neutron_SU3.eps}}
\hspace{8pt}
\subfloat[]{
\label{fig:KostasplotsNLO:f} 
\includegraphics[width=0.45\linewidth]{Kaon_Xi_SU3.eps}}
\caption{Plots of $\Gamma_{NLO}$ and $\Gamma_{NNLO}$ versus the
  Goldstone masses.  The line at $\Gamma=1$ is the leading
  order curve, and dotted line is the physical meson mass. The
  innermost error bar is the statistical uncertainty and the outermost error
  bar is the statistical and systematic uncertainty added in quadrature. The
  inner and outer filled bands correspond to the statistical and systematic uncertainty, respectively, of the fits to the LECs at NLO and NNLO
  using $\pi^+\Sigma^+$, and~$\pi^+\Xi^0$ {\it only}, for the SU(3) case.}
\label{fig:KostasNLONNLO}
\end{figure}

In the three-flavor chiral expansion, we have an overdetermined system
at both NLO and NNLO. While there are five observables, there are two
Low Energy Constants (LECs) at NLO, $C_0$ and $C_{01}$, and two LECs
at NNLO, $h_1$ and $h_{123}$. Fits of the LECs from each process at
NLO are given in Table~\ref{tab:LECpi} and the corresponding values of
the scattering lengths are given in Table~\ref{tab:scattLpisig}. At
NLO, the LECs are of natural size, and provide a consistent extraction within
uncertainties. Correspondingly, the scattering lengths appear to deviate
perturbatively from the LO values. The perturbative behavior of the
scattering lengths at NLO is evident from the plots of $\Gamma_{NLO}$
versus the Goldstone masses given in Fig.~\ref{fig:KostasNLONNLO}.
Clearly the deviations of the lattice data from unity are consistent
with a perturbative expansion.

At NNLO the situation changes dramatically. This is clear from the
plots of $\Gamma_{NNLO}$ versus the Goldstone masses given in
Fig.~\ref{fig:KostasNLONNLO}. The shift of the value of $\Gamma$
from NLO to NNLO is dependent on the renormalization scale $\mu$.
With the choice $\mu=\Lambda_\chi$ one would expect this shift to be
perturbative. However, this is not the case and therefore loop
corrections are very large at the scale $\Lambda_\chi$. There are many
strategies that one may take to fit the LECs in the overdetermined
system.  Here we fit the LECs to the $\pi^+\Sigma^+$ and $\pi^+\Xi^0$
data, and then use these LECs to predict the kaon
processes. Therefore, in Fig.~\ref{fig:KostasNLONNLO}, only (a) and
(b) are fits.  The fit LECs are given in Table~\ref{tab:LECpi}.  While
the NNLO LECs $h_1$ and $h_{123}$ appear to be of natural size, the
NLO LECs $C_0$ and $C_{01}$ are unnaturally large and therefore are
countering the large loop effects. The extrapolated $\pi^+\Sigma^+$
and $\pi^+\Xi^0$ scattering lengths are given in
Table~\ref{tab:scattLpisig} and appear to be
perturbative. Table~\ref{tab:scattLpisig} also gives the extrapolated
kaon-baryon scattering lengths with the LECs determined from the
$\pi^+\Sigma^+$ and $\pi^+\Xi^0$ data. The resulting NNLO predictions
deviate by at least 100\% from the LO values.  Other fitting
strategies lead to this same conclusion: the kaon-baryon scattering
lengths are unstable against chiral corrections in the three-flavor
chiral expansion, over the range of light-quark masses that we
consider.

\begin{table}
\begin{ruledtabular}
\begin{tabular}{cccc}
Quantity & NLO fit each process &  NNLO fit $\pi^+\Sigma^+$,$\pi^+\Xi^0$  \\
\hline
$C_1(\pi^+\Sigma^+)$ &  0.66(04)(11) GeV$^{-1}$ & 3.51(18)(25) GeV$^{-1}$ \\
$C_{01}(\pi^+\Xi^0)$ &  0.69(06)(22) GeV$^{-1}$ & 7.44(29)(69) GeV$^{-1}$ \\
$C_1(K^+ p)$ &  0.44(09)(23) GeV$^{-1}$ & - \\
$C_{01}(K^+ n)$ &  0.56(11)(27) GeV$^{-1}$ & - \\
$C_1(\overline{K}{}^0\Xi^0)$ &  0.50(06)(14) GeV$^{-1}$ & - \\
\hline
$h_1$ & - & -0.59(08)(14) GeV$^{-2}$  \\
$h_{123}$ & - & -0.42(10)(10) GeV$^{-2}$ \\
\end{tabular}
\end{ruledtabular}
\caption{$SU(3)$ LECs fit from each process at NLO, and from $\pi^+\Sigma^+$, and~$\pi^+\Xi^0$ at NNLO. 
The first uncertainty in parentheses is statistical, and the second is the statistical and systematic uncertainty added in quadrature.}
\label{tab:LECpi}
\end{table}
\begin{table}
\begin{ruledtabular}
\begin{tabular}{ccccc}
Quantity & LO (fm) & NLO fit (fm) & NLO (NNLO fit) (fm) & NNLO (fm) \\
\hline
$a_{\pi\Sigma}$ & -0.2294 & -0.208(01)(03) & -0.117(06)(08) & -0.197(06)(08) \\
$a_{\pi\Xi}$ & -0.1158 & -0.105(01)(04) &  0.004(05)(11) & -0.096(05)(12) \\
$a_{Kp}$ & -0.3971 & -0.311(18)(44) &  0.292(35)(48) & -0.154(51)(63) \\
$a_{Kn}$ & -0.1986 & -0.143(10)(27) &  0.531(28)(68) &  0.128(42)(87) \\
$a_{K\Xi}$ & -0.4406 & -0.331(12)(31) &  0.324(39)(54) & -0.127(57)(70) \\
\end{tabular}
\end{ruledtabular}
\caption{$SU(3)$ extrapolated scattering lengths using the LECs from
  Table~\ref{tab:LECpi}. The first uncertainty in parentheses is
  statistical, and the second is the statistical and systematic uncertainty
  added in quadrature. Note that the NLO (NNLO fit) column is using $C_1,C_{01}$
  from the NNLO fit to $\pi^+\Sigma^+$,$\pi^+\Xi^0$.}
\label{tab:scattLpisig}
\end{table}

\section{SU(2) HB$\chi PT$ Extrapolation}
\label{sec:su2CE}

\noindent Given the poor convergence seen in the three-flavor chiral
expansion due to the large loop corrections, it is natural to consider
the two-flavor theory with the strange quark integrated out. In this
way, $\pi\Sigma$ and $\pi\Xi$ may be analyzed in an expansion in
$m_\pi$ with no fear of corrections that scale as powers of $m_K$.
The detailed matching of LECs between the three- and two-flavor
theories is described in detail in Ref.~\cite{Mai:2009ce}. We make use
of the formulation of the $\pi\Sigma$ and $\pi\Xi$ T-matrices
from~\cite{Mai:2009ce} to perform the two-flavor chiral extrapolations
for $a_{\pi^+\Sigma^+}$, and $a_{\pi^+\Xi^0}$. As pointed out in
Ref.~\cite{Mai:2009ce}, there are two representations of the
pion-hyperon scattering lengths that are equivalent up to omitted
higher orders in the chiral expansion; one contains a chiral
logarithm, and the other is purely a polynomial in $m_\pi$. Using both
forms provides a useful check on the systematics of the chiral
extrapolation.

\subsection{Scattering Length Formulas I}

\begin{figure}[ht]
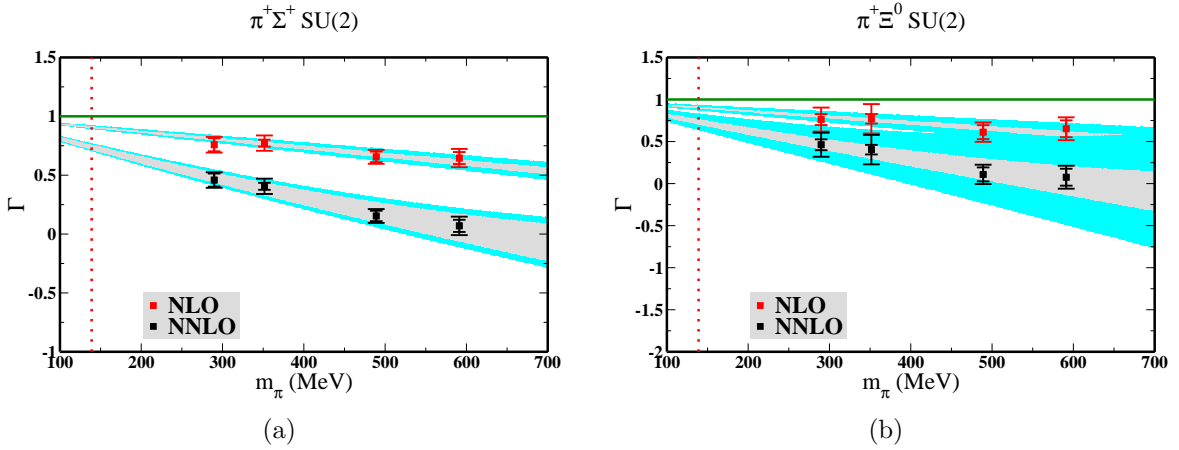

\centering
\subfloat[]{
\label{fig:KostasSU2:a} 
\includegraphics[width=0.45\linewidth]{Pi_Sigma_SU2.eps}}
\hspace{8pt}
\vspace{10pt}
\subfloat[]{
\label{fig:KostasSU2:b} 
\includegraphics[width=0.45\linewidth]{Pi_Xi_SU2.eps}}
\caption{$\Gamma_{NLO}$, $\Gamma_{NNLO}$ plots for the
  $\pi^+\Sigma^+$, and~$\pi^+\Xi^0$ processes versus the pion
  mass. The line at $\Gamma=1$ is the leading order
  curve, and the dotted line is the physical pion mass. The
  innermost error bar is the statistical uncertainty and the outermost error
  bar is the statistical and systematic uncertainty added in quadrature. The
  inner and outer filled bands correspond to the statistical and systematic uncertainty, respectively, of the fits to the LECs at NLO and NNLO
  using $\pi^+\Sigma^+$, and~$\pi^+\Xi^0$ for the SU(2) case.}
\label{fig:KostasSU2}
\end{figure}

\begin{figure}[h]
\centering
\includegraphics[width=0.49\linewidth]{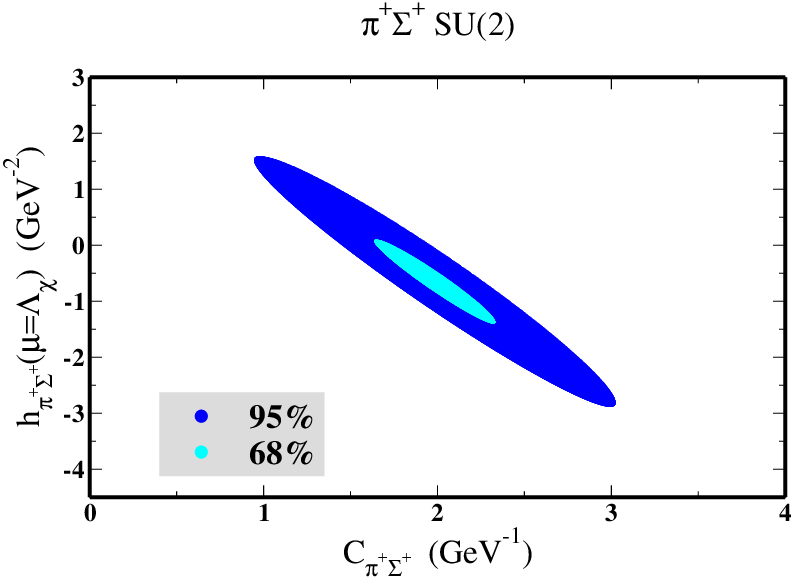}\hfill
\includegraphics[width=0.49\linewidth]{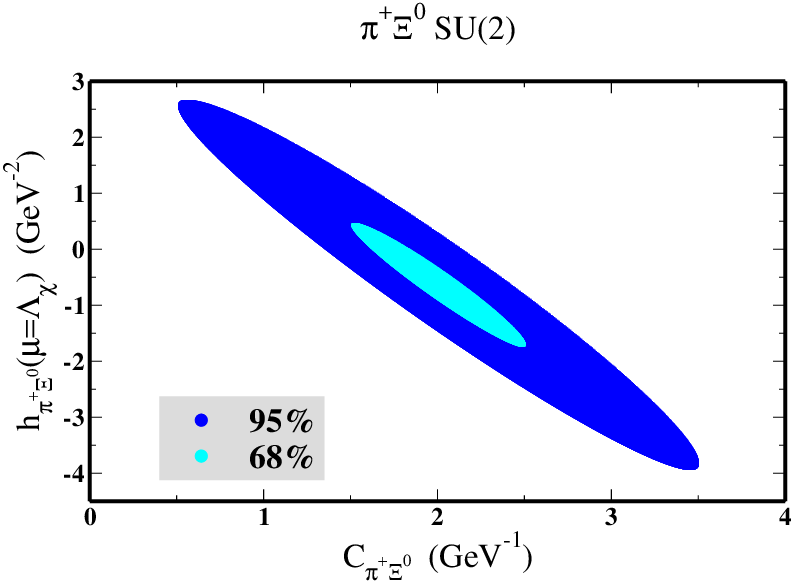}\hfill
\caption{The 68\% (light) and 95\% (dark) confidence interval error 
ellipses for fits for the $\pi^+\Sigma^+$ (left), and~$\pi^+\Xi^0$ (right) processes
using Eqs.~\protect(\ref{eq:apisigSU2}) and \protect(\ref{eq:apixiSU2}).}
\label{fig:errellipseK} 
\end{figure}

\noindent To $\mathcal{O}(m_\pi^3)$ in the two-flavor chiral expansion, $a_{\pi^+\Sigma^+}$ and $a_{\pi^+\Xi^0}$ are given by~\cite{Mai:2009ce}
\begin{eqnarray}
a_{\pi^+\Sigma^+}=\frac{1}{4 \pi}\frac{m_\Sigma}{m_\pi+m_\Sigma} \bigg[ -\frac{2m_\pi}{f_\pi^2} + \frac{2m_\pi^2}{f_\pi^2} {\mathrm{C}}_{\pi^+\Sigma^+} 
+\frac{m_\pi^3}{\pi^2 f_\pi^4}\log{\frac{m_\pi}{\mu}} + \frac{2m_\pi^3}{f_\pi^2}{h}_{\pi^+\Sigma^+}(\mu ) \bigg] \ ;
\label{eq:apisigSU2}
\end{eqnarray}

\begin{eqnarray}
a_{\pi^+\Xi^0}=\frac{1}{4 \pi}\frac{m_\Xi}{m_\pi+m_\Xi} \bigg[ -\frac{m_\pi}{f_\pi^2} + \frac{m_\pi^2}{f_\pi^2}{\mathrm{C}}_{\pi^+\Xi^0} 
+ \frac{m_\pi^3}{2\pi^2 f_\pi^4}\log{\frac{m_\pi}{\mu}} + \frac{m_\pi^3}{f_\pi^2}{h}_{\pi^+\Xi^0}(\mu ) \bigg]
\label{eq:apixiSU2} \ ,
\end{eqnarray}
where the explicit forms ---in terms of Lagrangian parameters--- of the LECs
${\mathrm{C}}_{\pi^+\Sigma^+}$, ${h}_{\pi^+\Sigma^+}$,
${\mathrm{C}}_{\pi^+\Xi^0}$ and ${h}_{\pi^+\Xi^0}$ are given in
Ref.~\cite{Mai:2009ce}. As in the three flavor case, the mixed-action modification to the  
$SU(2)$ scattering length formula would begin with corrections to the  
$m_\pi^3 \ln (m_\pi)$ terms, with the mixed valence-sea pions having  
the known additive mass shift~\cite{Orginos:2007tw}. We again choose $\mu=\Lambda_\chi=4\pi f_\pi$
and evaluate $f_\pi$ at its lattice physical value.
In analogy with the three-flavor case, we define
\begin{eqnarray}
\Gamma_{LO}\equiv 1 \ ;
\label{eq:GammaLHSLO1su2}
\end{eqnarray}
\begin{eqnarray}
\Gamma_{NLO}\equiv 1-C_{\pi^+ B} m_\pi \ ;
\label{eq:GammaLHSNLO1su2}
\end{eqnarray}
\begin{eqnarray}
\Gamma_{NNLO}\equiv
1-C_{\pi^+ B} m_\pi-h_{\pi^+ B}(\Lambda_{\chi}) m_\pi^2 \ ,
\label{eq:GammaLHSNNLO1su2}
\end{eqnarray} 
where $B$ is either $\Sigma^+$ or $\Xi^0$.  In
Fig.~\ref{fig:KostasSU2} we give plots of $\Gamma_{NLO}$ and
$\Gamma_{NNLO}$ versus the pion mass for the two-flavor
case. Clearly the deviations of $\Gamma$ from unity are
consistent with a perturbative expansion at both NLO and NNLO, showing that the loop corrections are much smaller at the scale $\Lambda_\chi$ than in the three-flavor case. All extracted LECs are of natural size and given in
Table~\ref{tab:LECpisu2I}. The extrapolated $\pi^+\Sigma^+$ and $\pi^+\Xi^0$ scattering
lengths are given in Table~\ref{tab:scattLpisigsu2I}. The results are
consistent with what was found in the three-flavor extrapolation.  The
NLO and NNLO LECs are highly correlated in the NNLO fit.
Fig.~\ref{fig:errellipseK} shows the 68\% and 95\% confidence
interval error ellipses in the $h$-${\mathrm{C}}$ plane for both
${\pi^+\Sigma^+}$ and ${\pi^+\Xi^0}$.  Exploring the full 95\%
confidence interval error ellipse in the $h$-${\mathrm{C}}$ plane
yields
\begin{eqnarray}
a_{\pi^+\Sigma^+}&=& -0.197 \pm 0.017~{\rm fm} \ ;\\
a_{\pi^+\Xi^0}&=& -0.098\pm 0.017~{\rm fm} \ .
\label{eq:MP2}
\end{eqnarray}
These are the numbers that we quote as our best determinations of the pion-hyperon
scattering lengths.

\begin{table}
\begin{center}
\vskip 0.2cm
\resizebox{8cm}{!}
{
\begin{tabular}{ccc}
\hline
\hline
 & NLO fit & NNLO fit \\
\hline
${C}_{\pi^+\Sigma^+}$ &  0.66(04)(11) GeV$^{-1}$ &  1.98(17)(24) GeV$^{-1}$ \\
${C}_{\pi^+\Xi^0}$ &  0.69(06)(22) GeV$^{-1}$ &  2.01(24)(68) GeV$^{-1}$ \\
\hline
$h_{\pi^+\Sigma^+}$ & - & -0.65(36)(40) GeV$^{-2}$ \\
$h_{\pi^+\Xi^0}$ & - & -0.6(0.5)(1.1) GeV$^{-2}$ \\
\hline
\hline
\end{tabular}
}
\caption{$SU(2)$ LECs fit from each process at NLO and at NNLO.  The
  first uncertainty in parentheses is statistical, and the second is the
  statistical and systematic uncertainty added in quadrature.}
\label{tab:LECpisu2I}
\end{center}
\end{table}
\begin{table}
\begin{center}
\vskip 0.2cm
\resizebox{12cm}{!}
{
\begin{tabular}{ccccc}
\hline
\hline
Quantity & LO (fm) & NLO (fm) & NLO (NNLO fit) (fm) & NNLO (fm) \\
\hline
$a_{\pi\Sigma}$ & -0.2294 & -0.208(01)(03) & -0.166(05)(08) & -0.197(06)(08) \\
$a_{\pi\Xi}$ & -0.1158 & -0.105(01)(04) & -0.083(04)(11) & -0.098(05)(12) \\
\hline
\hline
\end{tabular}
}
\caption{$SU(2)$ extrapolated scattering lengths using the LECs from
  Table~\ref{tab:LECpisu2I}. The first uncertainty in parentheses is
  statistical, and the second is the statistical and systematic uncertainty
  added in quadrature.}
\label{tab:scattLpisigsu2I}
\end{center}
\end{table}

\subsection{Scattering Length Formulas II}

\begin{figure}[!h]
\centering
\includegraphics[width=0.49\linewidth]{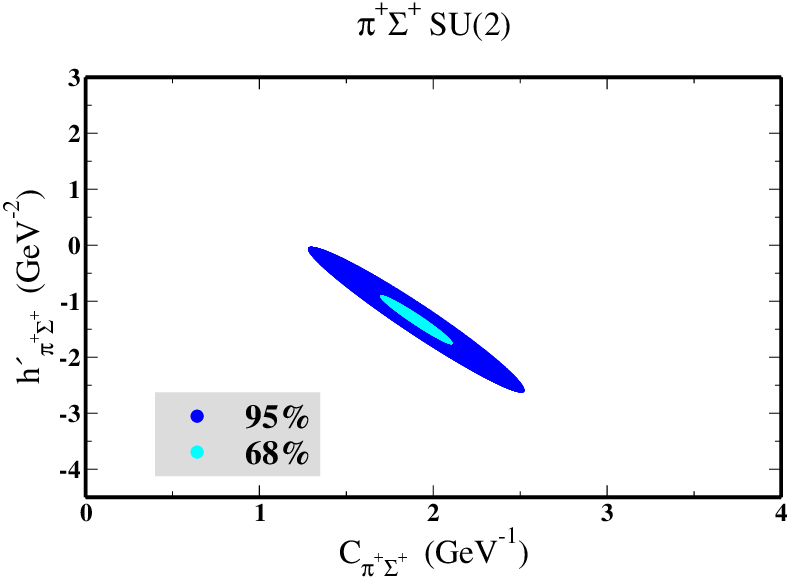}\hfill
\includegraphics[width=0.49\linewidth]{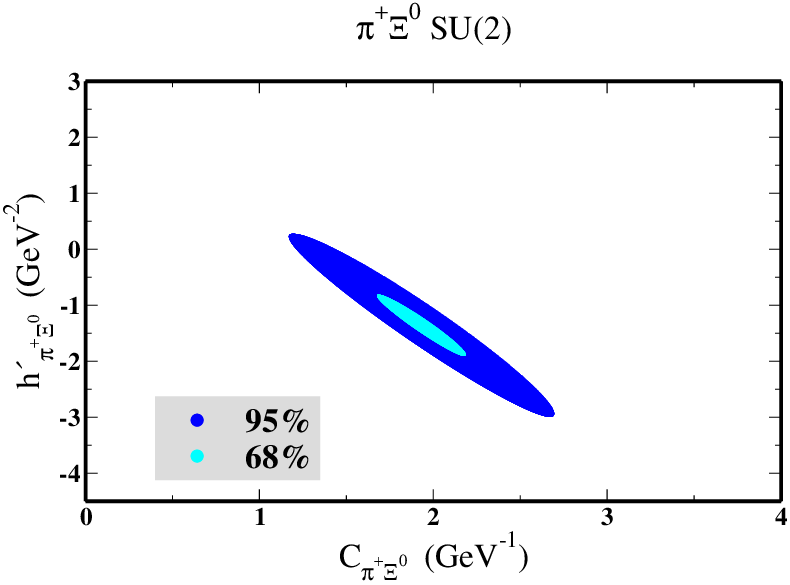}\hfill
\caption{The 68\% (light) and 95\% (dark) confidence interval error 
ellipses for fits for the $\pi^+\Sigma^+$ (left), and~$\pi^+\Xi^0$ (right) processes
using Eqs.~\protect(\ref{eq:apisig2param}) and \protect(\ref{eq:apixi2param}).}
\label{fig:errellipseU} 
\end{figure}

\noindent Ref.~\cite{Mai:2009ce} makes the interesting observation that replacing $f_\pi$ with its chiral limit value,
$f$, yields

\begin{eqnarray}
a_{\pi^+\Sigma^+}=\frac{1}{2 \pi}\frac{m_\Sigma}{m_\pi+m_\Sigma} \bigg[ -\frac{m_\pi}{f^2} + \frac{m_\pi^2}{f^2} {\mathrm{C}}_{\pi^+\Sigma^+} 
+ \frac{m_\pi^3}{f^2} h'_{\pi^+\Sigma^+} \bigg], \qquad
h'_{\pi^+\Sigma^+}=\frac{4}{f^2}\ell_4^r+ h_{\pi^+\Sigma^+} \ ;
\label{eq:apisig2param}
\end{eqnarray}
\begin{eqnarray}
a_{\pi^+\Xi^0}=\frac{1}{4 \pi}\frac{m_\Xi}{m_\pi+m_\Xi} \bigg[ -\frac{m_\pi}{f^2} + \frac{m_\pi^2}{f^2}{\mathrm{C}}_{\pi^+\Xi^0} 
+ \frac{m_\pi^3}{f^2} h'_{\pi^+\Xi^0} \bigg],\qquad
h'_{\pi^+\Xi^0}=\frac{4}{f^2}\ell_4^r + h_{\pi^+\Xi^0} \ ,
\label{eq:apixi2param}
\end{eqnarray}
where $\ell_4^r$ is the LEC which governs the pion mass
dependence of $f_\pi$~\cite{Colangelo:2001df}.  Note that the chiral
logs have canceled, and in this form, valid to order
$m_\pi^3$ in the chiral expansion, the scattering lengths have a
simple polynomial dependence on $m_\pi$.  Taking the standard value
$f=122.9$ MeV~\cite{Colangelo:2001df,Mai:2009ce} and refitting the LECs
yields the results tabulated in
Table~\ref{tab:LECpisu2II}. The extrapolated $\pi^+\Sigma^+$ and $\pi^+\Xi^0$ scattering
lengths are given in Table~\ref{tab:scattLpisigsu2II}. These results are
clearly consistent with what was found in the two-flavor extrapolation with
the chiral logarithm explicit.
Fig.~\ref{fig:errellipseU} shows the 68\% and 95\% confidence
interval error ellipses in the $h$-${\mathrm{C}}$ plane for both
${\pi^+\Sigma^+}$ and ${\pi^+\Xi^0}$.  Exploring the full 95\%
confidence interval error ellipse in the $h$-${\mathrm{C}}$ plane
yields
\begin{eqnarray}
a_{\pi^+\Sigma^+}&=& -0.197 \pm 0.011~{\rm fm} \ ;\\
a_{\pi^+\Xi^0}&=& -0.102 \pm 0.004~{\rm fm} \ .
\label{eq:MPU}
\end{eqnarray}
Comparison of these determinations with those of Eq.~(\ref{eq:MP2}) give
an estimate of the systematic error due to truncation of the chiral
expansion at order $m_\pi^3$. We have also ``pruned'' the data; that is,
we have redone all fits omitting the heaviest mass ensemble. While this
procedure inflates the errors, we see very little shift in the central
values.
\begin{table}
\begin{center}
\vskip 0.2cm
\resizebox{8cm}{!}
{
\begin{tabular}{ccc}
\hline
\hline
 & NLO fit & NNLO fit \\
\hline
$C_{\pi^+\Sigma^+}$ &  1.28(09)(11) GeV$^{-1}$ &  1.90(10)(17) GeV$^{-1}$ \\
$C_{\pi^+\Xi^0}$ &  1.84(23)(25) GeV$^{-1}$ &  1.93(12)(48) GeV$^{-1}$ \\
\hline
$h^{'}_{\pi^+\Sigma^+}$ & - & -1.33(21)(26) GeV$^{-2}$ \\
$h^{'}_{\pi^+\Xi^0}$ & - & -1.36(27)(75) GeV$^{-2}$ \\
\hline
\hline
\end{tabular}
}
\caption{$SU(2)$ LECs fit from each process at NLO and at NNLO.  The
  first uncertainty in parentheses is statistical, and the second is the
  statistical and systematic uncertainty added in quadrature.}
\label{tab:LECpisu2II}
\end{center}
\end{table}

\begin{table}
\begin{center}
\vskip 0.2cm
\resizebox{12cm}{!}
{
\begin{tabular}{ccccc}
\hline
\hline
Quantity & LO (fm) & NLO (fm) & NLO (NNLO fit) (fm) & NNLO (fm) \\
\hline
$a_{\pi\Sigma}$ & -0.2294 & -0.212(03)(04) & -0.190(04)(06) & -0.197(04)(09) \\
$a_{\pi\Xi}$ & -0.1158 & -0.106(04)(05) & -0.095(02)(09) & -0.102(02)(09) \\
\hline
\hline
\end{tabular}
}
\caption{$SU(2)$ extrapolated scattering lengths using the LECs from
  Table~\ref{tab:LECpisu2II}. The first uncertainty in parentheses is
  statistical, and the second is the statistical and systematic uncertainty
  added in quadrature.}
\label{tab:scattLpisigsu2II}
\end{center}
\end{table}

\begin{figure}[ht]
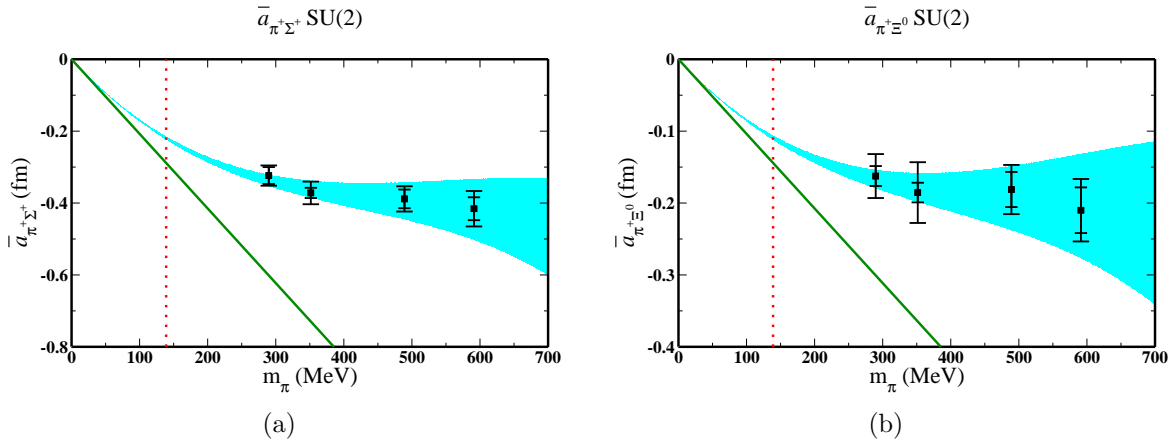

\centering
\subfloat[]{
\label{fig:aSU2:a} 
\includegraphics[width=0.45\linewidth]{a_Pi_Sigma_SU2.eps}}
\hspace{8pt}
\vspace{10pt}
\subfloat[]{
\label{fig:aSU2:b} 
\includegraphics[width=0.45\linewidth]{a_Pi_Xi_SU2.eps}}
\caption{$\overline{a}$ plots for the $\pi^+\Sigma^+$,
  and~$\pi^+\Xi^0$ processes versus the pion mass. The
  diagonal line is the leading order curve, and the dotted line is the
  physical pion mass. The innermost error bar is the statistical uncertainty
  and the outermost error bar is the statistical and systematic uncertainty
  added in quadrature. The filled bands are the fits to the LECs in the
  SU(2) case at NNLO as in Eqs.~(\ref{eq:abarpisigSU2}), and~(\ref{eq:abarpixiSU2}).}
\label{fig:aSU2}
\end{figure}

In order to plot the scattering length versus $m_\pi$, we define 
\begin{eqnarray}
\overline{a}_{\pi^+\Sigma^+}=a_{\pi^+\Sigma^+}\left(\frac{m_\pi+m_\Sigma}{m_\Sigma} \right) =\frac{1}{2\pi}\left( -\frac{m_\pi}{f^2} + \frac{m_\pi^2}{f^2} {\mathrm{C}}_{\pi^+\Sigma^+} + \frac{m_\pi^3}{f^2} h'_{\pi^+\Sigma^+} \right) \ ;
\label{eq:abarpisigSU2}
\end{eqnarray}
\begin{eqnarray}
\overline{a}_{\pi^+\Xi^0}=a_{\pi^+\Xi^0}\left(\frac{m_\pi+m_\Xi}{m_\Xi} \right)=\frac{1}{4\pi}\left( -\frac{m_\pi}{f^2} + \frac{m_\pi^2}{f^2}{\mathrm{C}}_{\pi^+\Xi^0} 
+ \frac{m_\pi^3}{f^2} h'_{\pi^+\Xi^0} \right) \ .
\label{eq:abarpixiSU2}
\end{eqnarray}
In Fig.~\ref{fig:aSU2} we plot the scattering lengths versus the pion mass. The shaded bands in these
plots correspond to the standard error in the determination of the LECs, as given in Table~\ref{tab:LECpisu2II}.

Additional systematic errors arising from the specific lattice formulation that we employ
are discussed in detail in Ref.~\cite{Beane:2007xs}, and are expected to be well encompassed
by our error bars. As discussed in section~\ref{sec:finvol}, there is a systematic error in
extracting the scattering length from the phase shift. We find that range corrections
affect the scattering length at the 5\% level for $\pi^+\Sigma^+$, and at the 1\%
level for $\pi^+\Xi^0$. Finally, we reiterate that there are unquantified systematic errors due to finite-volume and lattice-spacing effects, however, these errors are likely encompassed by our quoted errors.

\section{Conclusions}
\label{sec:conc}

\noindent In this paper we have presented the first fully-dynamical
lattice QCD calculation of meson-baryon scattering. While the
phenomenologically most-interesting case of pion-nucleon scattering
involves annihilation diagrams, and therefore, requires more resources
than we currently have available, we have calculated the ground-state
energies of $\pi^+\Sigma^+$, $\pi^+\Xi^0$, $K^+p$, $K^+n$, and
$\overline{K}{}^0 \Xi^0$, which involve no annihilation diagrams.

An analysis of the scattering lengths of these two-body systems using
HB$\chi$PT has led us to conclude that the three-flavor chiral
expansion does not converge over the range of light quark masses that
we investigate.  While the kaon-baryon scattering lengths appear
perturbative at NLO, a comparison of NNLO with NLO calls into question
the convergence of the three-flavor chiral expansion. Therefore, we do
not quote values for the kaon-baryon scattering lengths at the
physical point. On the other hand, the $\pi^+\Sigma^+$ and
$\pi^+\Xi^0$ scattering lengths appear to have a well-controlled
chiral expansion in two-flavor HB$\chi$PT. Our results,
$a_{\pi^+\Sigma^+}=-0.197\pm0.017$ fm, and
$a_{\pi^+\Xi^0}=-0.098\pm0.017$ fm, deviate from the LO (current
algebra) predictions at the one- and two-sigma level, respectively. We
look forward to confirmation of these predictions from other lattice
QCD calculations and possibly from future experiments.

The HB$\chi$PT analyses performed in this work support a general
observation about convergence in the three-flavor chiral expansion, at
least for the processes studied here. As the pion masses considered in
this lattice calculation are comparable to the physical kaon mass, the
distinct convergence patterns of the two- and three-flavor chiral
expansions found in this work are suggestive that the breakdown in the
three-flavor case is not due to the relative largeness of the
strange-quark mass as compared to the light quark masses, but rather
due to some other enhancement in the coefficients of the loop
contributions, possibly related to a scaling with powers of $n_f$, the
number of flavors.

While in this paper we have not considered the lowest-lying baryon
decuplet, one interesting process for future study is the
$\pi^-\Omega^-$ system. It does not involve disconnected diagrams
since the pions have no valence quarks with the same flavor as the
$\Omega^-$ constituents. It has been argued that there is a bound
state~\cite{Wang:2006jg} in this channel, and therefore, it would be of
interest to determine whether this state appears bound on the lattice
at the available quark masses.

\section{Acknowledgments}

\noindent 
We thank U.G.~Mei\ss ner for useful discussions, and R.~Edwards and
B.~Joo for help with the QDP++/Chroma programming
environment~\cite{Edwards:2004sx} with which the calculations
discussed here were performed.  We gratefully acknowledge the
computational time provided by NERSC (Office of Science of the
U.S. Department of Energy, No. DE-AC02-05CH11231), the Institute for
Nuclear Theory, Centro Nacional de Supercomputaci\'on (Barcelona,
Spain), Lawrence Livermore National Laboratory, and the National
Science Foundation through Teragrid resources provided by the National
Center for Supercomputing Applications and the Texas Advanced
Computing Center.  Computational support at Thomas Jefferson National
Accelerator Facility and Fermi National Accelerator Laboratory was
provided by the USQCD collaboration under {\it The Secret Life of a
  Quark}, a U.S. Department of Energy SciDAC project ({\tt
  http://www.scidac.gov/physics/quarks.html}).  The work of MJS was
supported in part by the U.S.~Dept.~of Energy under Grant
No.~DE-FG03-97ER4014.  The work of KO and WD was supported in part by
the U.S.~Dept.~of Energy contract No.~DE-AC05-06OR23177 (JSA) and DOE
grant DE-FG02-04ER41302. KO and AWL were supported in part by the
Jeffress Memorial Trust, grant J-813 and DOE OJI grant
DE-FG02-07ER41527. The work of SRB and AT was supported in part by the
National Science Foundation CAREER grant No.  PHY-0645570.  Part of
this work was performed under the auspices of the US DOE by the
University of California, Lawrence Livermore National Laboratory under
Contract No. W-7405-Eng-48. The work of AP was partly supported by the
EU contract FLAVIAnet MRTN-CT-2006-035482, by the contract
FIS2008-01661 from MEC (Spain) and FEDER and by the Generalitat de
Catalunya contract 2005SGR-00343.

%
%

%
%

\end{document}